\documentclass[twocolumn]{article}

\usepackage[margin=0.5in]{geometry}

\usepackage[latin1]{inputenc}
\usepackage{amsmath, amstext, amsfonts}
\usepackage{graphicx}
\usepackage{array, color, colortbl}

\makeatletter
\let\@tabclassz =\@classz
\makeatother 

\usepackage{cite}
\usepackage{todonotes}
\usepackage{url}

\newcommand{\ve}[1]{\mathbf{#1}}

\newcommand{\T}{{\textsf{T}\!}}

\newcommand{\pfrst}{{\normalfont\textsc{p4est}}}

\newcommand{\aspect}{\textsc{Aspect}}

\definecolor{Gray}{gray}{0.9}

\begin{document}

%short title in []...
\title%[Modern numerical methods for mantle convection II: Realistic
  %models]
  {High Accuracy Mantle Convection Simulation through Modern Numerical
    Methods. II: Realistic Models and Problems}
% \author[T. Heister, J. Dannberg, R. Gassm{\"o}ller, W. Bangerth]
% {
% Timo Heister$^1$
% \and
% Juliane Dannberg$^{2}$\thanks{Now at Department of Mathematics, Colorado State University, Fort Collins, CO 80523-1874.}
% \and
% Rene Gassm{\"o}ller$^3$
% \and
% Wolfgang Bangerth$^4$ \\
% %
% \small $^{1}$ \it Mathematical Sciences, Clemson University, O-110 Martin Hall, Clemson, SC 29634-0975, USA; {\tt
%   heister@clemson.edu} \\
% \small $^{2}$ \it Department of Mathematics, Texas A\&M University, Mailstop 3368, College Station, TX 77843-3368, USA; {\tt
%   dannberg@math.tamu.edu} \\
% \small $^{3}$ \it Department of Mathematics,
% Colorado State University, Fort Collins, CO 80523-1874, USA; 
% {\tt rene.gassmoeller@mailbox.org} \\
% \small $^{4}$ \it Department of Mathematics,
% Colorado State University, Fort Collins, CO 80523-1874; 
% {\tt bangerth@colostate.edu}
% }

\author{
 Timo Heister$^1$
 \and
 Juliane Dannberg$^{2}$\thanks{Now at Department of Mathematics, Colorado State University, Fort Collins, CO 80523-1874.}
 \and
 Rene Gassm{\"o}ller$^3$
 \and
 Wolfgang Bangerth$^4$ \\
 \small $^{1}$ \it Mathematical Sciences, Clemson University, O-110 Martin Hall, Clemson, SC 29634-0975, USA; {\tt
   heister@clemson.edu} \\
 \small $^{2}$ \it Department of Mathematics, Texas A\&M University, Mailstop 3368, College Station, TX 77843-3368, USA; {\tt
   dannberg@math.tamu.edu} \\
 \small $^{3}$ \it Department of Mathematics,
 Colorado State University, Fort Collins, CO 80523-1874, USA; 
 {\tt rene.gassmoeller@mailbox.org} \\
 \small $^{4}$ \it Department of Mathematics,
 Colorado State University, Fort Collins, CO 80523-1874; 
 {\tt bangerth@colostate.edu}
}

\maketitle

NOTE: This paper has been published in Geophysical Journal International with
the title ``High Accuracy Mantle Convection Simulation through Modern Numerical Methods. II: Realistic Models and Problems'' with identifier \url{https://dx.doi.org/10.1093/gji/ggx195}.

\begin{abstract}
  Computations have helped elucidate the dynamics of Earth's mantle
  for several decades already. The numerical methods that underlie these simulations
  have greatly evolved within this time span, and today include dynamically
  changing and adaptively refined meshes, sophisticated and efficient solvers,
  and parallelization to large clusters of computers. At the same time, many
  of these methods -- discussed in detail in a previous paper in this series
  \cite{KHB12} -- were developed and tested primarily using model problems
  that lack many of the complexities that are common to the realistic models
  our community wants to solve today. 

  With several years of experience solving complex and realistic models, we
  here revisit some of the algorithm designs of the earlier paper and discuss
  the incorporation of more complex physics. In particular, we re-consider
  time stepping and mesh refinement algorithms, evaluate approaches to incorporate compressibility, and
  discuss dealing with strongly varying material coefficients, latent heat,
  and how to track chemical compositions and heterogeneities. Taken
  together and implemented in a high-performance, massively parallel code,
  the techniques discussed in this paper then allow for high resolution, 3d,
  compressible, global mantle convection simulations  
  with phase transitions, strongly temperature dependent viscosity and realistic material
  properties based on mineral physics data.

  Keywords:   Mantle convection, numerical methods, adaptive mesh refinement, finite
  element method, compressibility, preconditioners
  \end{abstract}
% \begin{keywords}
% Mantle convection, numerical methods, adaptive mesh refinement, finite
%   element method, compressibility, preconditioners
% \end{keywords}

\section{Introduction}

Computer simulations are at the heart of most attempts at understanding the
dynamics of the Earth's mantle as well as the interiors of other celestial bodies. As
such, there is a long tradition in the investigation of numerical methods that
help us solve the equations that describe mantle convection, dating back
many decades (e.g. \cite{torrance1971thermal,richter1973dynamical,mckenzie1974convection,
baumgardner1985three,tackley1993effects}, 
see also \cite{may2013overview} and references therein). 
Many of these articles parallel the general development of
computational science methods, and have moved from simple, low-order, uniform
2d mesh discretizations with fixed-point linear solvers, to using adaptively
refined, dynamically changing 3d meshes with higher order elements and
complicated linear and nonlinear solvers \cite{stadler2010dynamics,alisic2010slab,
davies2011fluidity,burstedde2013large,gerya2013adaptive,rudi2015extreme}. 
Indeed, a previous paper
\cite{KHB12} in the current series of publications
was devoted to the description of current, state-of-the-art
methods for mantle convection simulations.

At the same time, most of these methods -- including the ones in our earlier
paper -- were developed, tested, and evaluated using relatively simple model
problems (e.g. \cite{BBC89,BC93,keken1997comparison,TK03,Bab08,van2008community,
ZMTMG08,KLVLZTTK10,crameri2012comparison,tosi2015community}). 
Yet, this no longer matches what our community wants to do today:
We want to solve more realistic problems that use compressible
formulations with discontinuous coefficients, for
example. We also want to use more complex geometries, possibly varying
with time. And we may want to include other physical effects such as
latent heat, the transport of chemical inhomogeneities or tracking of
tensor quantities like finite strain.
For these kinds of applications, we have found that the
numerical
methods currently used in our community often perform worse than for the traditional
model problems and benchmarks.

The purpose of this paper is therefore to \textit{revisit the
  traditional choices of numerical methods for mantle convection in light of complex
  applications}. Specifically, we will consider how time stepping,
mesh refinement, formulations for compressible materials, and other
aspects of computational codes are affected when
they are applied to complex problems.
In some cases, previous methods perform poorly and need to be
adapted; in others, previous methods were simply unsuitable, and we
are faced with a variety of choices that allow us to design algorithms
that are both well suited to the problem as well as allow for accurate
and fast solutions.

We base our discussions on the five years of experience we
have with the \aspect{} code%
\footnote{The ``Advanced Solver for Problems in
Earth ConvecTion'', an open source project to provide a modern,
parallel, extensible code to simulate mantle convection. \aspect{}'s
development is supported by the Computational Infrastructure for
Geodynamics initiative, as well as by the National Science
Foundation. See \url{http://aspect.dealii.org}.}
since we described many of these methods
in \cite{KHB12}. In this time, we and others have applied \aspect{} to
more complex and realistic problems 
\cite{austermann2015impact,tosi2015community,rose2016stability,gassmoller2016major,dannberg2016compressible,zhang2016early,he2016discontinuous}, and the discussions in the remainder of this paper reflect the
challenges encountered in this process. On the other hand, the
discussions herein are not specific to \aspect{}: They are about the
general design of numerical methods for the problems at hand, and apply
equally to any other code that wants to solve them.

%\todo[inline]{other codes to look at:\\
% RHEA \cite{burstedde2013large}\\
% fluidity \cite{davies2011fluidity}\\
% terra \cite{baumgardner1985three}, 
% terra-neo\\
% citcom \cite{ZMTMG08} \\
% stagyy \cite{tackley2008modelling} \\
% lamem \\
% terraferma \cite{wilson2016terraferma} \\
% ConMan \\
% ASG stencil \cite{gerya2013adaptive}
%}

We intend this contribution to be of interest to those designing their
own numerical methods for mantle convection, but also for those
interested in understanding more about how modern mantle convection codes
work. Finally, some of the sections below outline open problems that
call for more methodological or mathematical research; the paper should
therefore also be of interest to the numerical methods and numerical analysis community
as it outlines areas requiring better methods.

The remainder of this paper is structured as follows:
Section~\ref{sec:formulation} first lays out the general mathematical
formulation of the problem we want to
consider. Section~\ref{sec:methods} then discusses how 
time stepping methods need to be adjusted to more complex problems
(Section~\ref{sec:timestepping});
how approaches can be designed to deal with compressibility
(Section~\ref{sec:compressibility}),
averaging discontinuous coefficients (Section~\ref{sec:averaging}),
and latent heat (Section~\ref{sec:latentheat});
how mesh refinement can be made to deal with realistic applications
(Section~\ref{sec:mesh-refinement});
and approaches to advecting along additional quantities
(Section~\ref{sec:composition}).
We show results for a large and complex application in
Section~\ref{sec:application}, and conclude in Section~\ref{sec:conclusions}.

\section{Formulation of the problem}
\label{sec:formulation}

Within this paper, let us consider a model for the flow of a compressible, anelastic fluid, 
such as generally assumed for convection in the Earth's mantle (e.g.~\cite{STO01}). 
Flow is driven by buoyancy due to thermal or compositional gradients,
and the model includes the effects of friction and adiabatic heating, radiogenic heat 
production and latent heat on the energy balance. 
However, the model ignores inertial and elastic
effects as we are concerned with very low velocities and long time scales.
Specifically, let us consider the following set of equations:
\begin{align}
  \label{eq:boussinesq-1}
  -\nabla \cdot \tau(\ve u) + \nabla p &= \rho \ve g,
  \\
  \label{eq:boussinesq-2}
  \nabla \cdot (\rho\ve u) &= 0,
  \\
  \rho C_p \left(
    \frac{\partial T}{\partial t} + \ve u \cdot \nabla T
  \right)
  - \nabla\cdot(k \nabla T)
  & = \rho H + \tau(\ve u) : \varepsilon(\ve u) \notag \\
  \label{eq:boussinesq-3}
  &\qquad  +\alpha T \left( \mathbf u \cdot \nabla p \right)
  \\
  &\qquad  + \rho T \frac{\mathrm{D} S}{\mathrm{D} t} \notag 
\end{align}
In this system of equations, $\ve u$ denotes the fluid velocity, $p$ the
pressure, and $T$ the temperature. For the stress we have
$\tau(\ve u) = 2\eta \left(\varepsilon(\ve u) - \frac 13 (\nabla \cdot \ve u)
  \ve I\right)$ with the rate-of-deformation tensor $\varepsilon (\ve
u)=\frac{1}{2} \left(\nabla \ve u+(\nabla \ve u)^\T\right)$.

In the equations above, $\eta,\rho$ and $C_p$ are the effective viscosity,
density, and specific heat capacity of the
material. $k,H,\alpha$, $\ve g$, and $S$ are the thermal conductivity,
intrinsic specific heat
production, thermal expansion coefficient, gravity vector, and entropy,
respectively. $\frac{\mathrm{D} S}{\mathrm{D} t}=\frac{\partial S}{\partial
  t}+\mathbf u \cdot \nabla S$ is the material derivative
of the entropy of a volume of material, and will be discussed in Section~\ref{sec:latentheat}.
We will in the following assume
that all of these parameters with the exception of gravity can depend on the
current temperature and
pressure; furthermore, we allow that $\eta$ can depend on the strain rate
$\varepsilon(\ve u)$ and that all parameters may also depend on the location
$\mathbf x$ to facilitate material parametrisations that are not derived from
realistic material models but incorporate a priori modeling assumptions. In
other words, we will henceforth consider
$\eta=\eta(p,T,\varepsilon(\ve u),\ve x)$,
$ \rho=\rho(p,T,\ve x)$,
$ \kappa=\kappa(p,T,\ve x)$,
$ H=H(p,T,\ve x)$,
$ \alpha=\alpha(p,T,\ve x)$,
$ \ve g =\ve g(\ve x) $.
Note that we assume the anelastic conservation of mass equation~\eqref{eq:boussinesq-2} and
only consider the density to be a dependent variable of temperature,
pressure, and location; in particular, we
neglect the time derivative and thus elastic waves, see \cite{STO01} for details.

In the remainder of this paper, we will make no
assumptions that coefficients are continuous. In fact, we explicitly allow
parameters to jump discontinuously as commonly happens when using
thermodynamically consistent models that incorporate phase changes. Indeed, it
is these kinds of difficulties that set apart the model problems often
considered, from the kind of problems that are the subject of this paper.

There are numerous approximations to equations
\eqref{eq:boussinesq-1}--\eqref{eq:boussinesq-3} that have been widely
used in the literature, such as the anelastic liquid approximation
(ALA), truncated anelastic liquid approximation (TALA) and Boussinesq 
approximation (BA), see for example \cite{BSG92, STO01, KLVLZTTK10, TG07}. 
These can all be derived by assuming that density variations are small 
compared to the hydrostatic density increase. We will discuss differences 
between these approximations and \eqref{eq:boussinesq-1}--\eqref{eq:boussinesq-3} 
in Section~\ref{sec:compressibility}, but these differences are not
fundamental to this paper: 
Any numerical issues that may arise from describing
the complex phenomena we aim to model would arise using any of the above formulations;
consequently, the solution strategies we derive are useful for all those cases.

\section{Numerical methods}
\label{sec:methods}

As discussed in the introduction, the goal of this section is to
outline areas where the numerical methods commonly employed for model
or simplified problems run into difficulties when applied to more
complex formulations and problems. The methods we compare against 
are Taylor-Hood finite elements to discretize the
Stokes equations, along with a block-preconditioned GMRES solver for
the resulting linear equations. The temperature equation is also
discretized using the finite element method; the advection is stabilized
via the addition of a nonlinear entropy viscosity. The entire set of
equation is discretized on adaptively refined, dynamically changing
meshes in 2d or 3d. All of these methods are
described in detail in a previous paper \cite{KHB12}. We consider them
state-of-the-art within the computational mantle convection community.

The focus of the following subsections is, then, on the modifications
necessary as one moves from simpler, model problems to the more
realistic description of convective transport in the Earth's mantle
provided in the previous section. Specifically, we will discuss time
stepping; dealing with compressibility; averaging of discontinuous
coefficients; incorporating latent heat; mesh adaptation; and
advection schemes for additional quantities.
On the other hand, we will not elaborate on the solution of models with non-linear, 
strain-rate dependent rheologies; a discussion and extensive benchmarking
of such cases can be found in \cite{Glerum2017}.

All computations in this section are done using the open source mantle convection code 
\aspect{} \cite{KHB12,aspectmanual}, which
builds on deal.II \cite{dealII84}, \pfrst{} \cite{p4est}, and Trilinos \cite{trilinos};
our test computations were done with \aspect{} version 1.5.0 \cite{aspect-doi-v1.5.0}
and the setups for all computations are available at
\url{https://github.com/tjhei/paper-aspect-methods-2-data}.

\subsection{Time stepping for the temperature equation revisited:
  explicit, semi-implicit, implicit}
\label{sec:timestepping}

In \cite{KHB12}, we have advocated for a semi-implicit method for the
time discretization of the temperature equation
\eqref{eq:boussinesq-3}. In this approach, one treats the thermal
diffusion term implicitly, but the advection term explicitly. This
choice guarantees that only the advection term implies a stability
limit for the size of the time step $\Delta t$. In particular, the
corresponding Courant-Friedrichs-Lewy (CFL) condition states that the
time discretized problem is only stable if we choose $\Delta t\le
\min_K C \frac{h_K}{p_T \|u\|_{L_\infty(K)}}$, where $h_K$ is a measure of
the (one-dimensional) size of cell $K$, $p_T$ the polynomial degree of the
finite element used to discretize the temperature field, and $\|u\|_{L_\infty(K)}$ is the maximal
velocity on cell $K$. $C$ is a constant related to the time stepping
method; it always satisfies $C\le 1$ if some terms of the equation are treated
explicitly, and generally becomes smaller with increasing convergence order of
the chosen time stepping method.

In pursuing this strategy, we were motivated by two observations. First,
the matrix that needs to be inverted in solving the temperature
equation with this choice of terms treated implicitly yields a
symmetric and positive definite matrix for which efficient solution
methods are readily available, in particular the Conjugate Gradient
method combined with multigrid preconditioners. Second, while fully implicit methods
may choose time steps much larger than $\min_K
\frac{h_K}{p_T \|u\|_{L_\infty(K)}}$ and still remain stable, we typically
  want to choose the time step around $\min_K
  \frac{h_K}{p_T \|u\|_{L_\infty(K)}}$ anyway for \textit{accuracy} reasons
    because this guarantees that information is not transported across
    more than the distance between adjacent nodes within one time step. This
    is of increasing importance when using adaptive mesh refinement.

However, in applying this approach to more realistic problems, one
encounters two difficulties:
\begin{itemize}
  \item How exactly should $h_K$ be defined?
  \item How large or small does one have to choose $C$?
\end{itemize}
These questions are relatively easy to answer on uniform meshes for
rectangular or box-shaped meshes. There, all possible definitions of $h_K$ --
either (i) the diameter of cell $K$, (ii) the shortest edge of $K$,
(iii) the minimal distance between any two vertices,
(iv) the square or cube root of the volume of the cell in 2d and 3d,
respectively -- are all equivalent up to a fixed constant and any
choice is valid as long as the constant $C$ is appropriately
adjusted. After choosing any of these definitions, we can determine
a safe value for $C$ experimentally.

On the other hand we desire to solve problems on complex domains that
will include cells of varying shapes; examples are meshes that
discretize models on shell segments, but also may have a free top
boundary and/or describe real topology. In such meshes, the various
ways of defining what $h_K$ is, are no longer equivalent up to a fixed
constant, and it is not clear what definition is the most appropriate
to allow for the largest choice of time steps.

Secondly, it may require extensive test simulations to determine whether a
particular choice of $C$ leads to a stable scheme because a solution
may only ``blow up'' once steep features of the solution happen to
pass a particularly poorly shaped cell, rather than such a steep
feature simply existing.

The consequence of all of this is that in practice, one needs to
choose $C$ rather small to guarantee stability in all
circumstances. As stated in \cite{KHB12}, we needed to choose
$C=\frac{1}{5.9}$ in 2d and $C=\frac{1}{43.6}$ in 3d. For any larger value, we could
find geometries and problem setups for which the temperature
eventually became instable, even though the resulting time steps are
almost certainly smaller than necessary for most other cases.

Such small time steps are impractical in practice. While the resulting
solution is stable, it is not significantly more accurate than if we
had chosen a fully implicit method with $\Delta t = \min_K \frac{h_K}{p_T \|u\|_{L_\infty(K)}}$.
However, using the semi-implicit scheme for the temperature equation is \textit{vastly more expensive}: we need
$\frac{1}{C}$ as many time steps for the semi-implicit method (i.e., roughly
one sixth of the number of time steps in 2d, and less than one fortieth in 3d), each
including solving both the Stokes and the temperature equation.

For these reasons, we have come to believe that the better choice for the time
stepping scheme is a fully implicit time
discretization -- for example a BDF-2 scheme to discretize the term
$\frac{\partial T}{\partial t}$ -- in which we choose $C=1$ and $h_K$ to be
the \textit{minimal distance between any two vertices} of cell $K$. Because
this choice treats advection
implicitly, it results in a system matrix that is no longer symmetric and positive
definite. This requires more costly
solvers and preconditioners, for example GMRES instead of CG. 
On the other hand, this effort is vastly
over-compensated by the fact that we have reduced the number of time
steps by a factor of more than 5 (in 2d) or 40 (in 3d). Furthermore,
even the fully implicit temperature solver requires less than 10\% of the overall run time
in realistic simulations; in other words, having to choose a less
efficient linear solver due to the addition of a non-symmetric term
does not affect the overall computational cost of each time step
in a significant way. What determines the overall computational cost of a
simulation, however, is how many time steps we have to solve.

\subsection{Compressibility}
\label{sec:compressibility}

Incorporating compressibility into existing codes is likely the most
difficult issue when moving from model problems to realistic
descriptions of Earth. This is because compressibility makes the mass
conservation equation \eqref{eq:boussinesq-2} nonlinear, or adds
additional terms when using the ALA or TALA
approximation. Furthermore, the divergence term is no longer adjoint
to the gradient of the pressure, and depending on how it is treated
numerically, the matrix resulting from the Stokes equation after
discretization may no longer be symmetric. As a consequence, how
exactly one deals with the compressibility has significant
implications for how nonlinear and linear solvers need to be written
and will perform. On the other hand, there are significant
opportunities for \textit{algorithm design} whereby one can
\textit{choose} different re-formulations based on which of these
allows for efficient and accurate implementations. The next
sub-section (Section~\ref{sec:compressible_stokes}) will therefore be
about the various trade-offs involved, before we comment on
considerations of the symmetry of resulting solvers
(Section~\ref{sec:compressible-symmetry}), making the right hand sides
of linear systems compatible (Section~\ref{sec:compressibility-rhs}),
and finally show numerical results illustrating several of the points
previously discussed (Sections~\ref{sec:tan_gurnis} and 
\ref{sec:compressible-2d-cartesian}).

Various forms of compressibility have been incorporated into mantle
convection codes for several decades already, though often only for
particular formulations such as the ALA or TALA in which the density
in the mass conservation equation is explicitly prescribed as a
function of depth. We refer to 
\cite{baumgardner1985three, TG07, LZ08, tackley2008modelling, KLVLZTTK10} 
for details on how other codes deal with these issues. 

\subsubsection{Reformulating the compressible Stokes equations}
\label{sec:compressible_stokes}

Solving compressible models numerically poses a number of challenges. First,
the mass conservation equation $\nabla \cdot (\rho\ve u)=0$ given
in \eqref{eq:boussinesq-2}, is nonlinear if $\rho$ depends on the
solution variables and has to be linearised. Secondly,
any linearised version results in an operator that is no longer
adjoint to the term $\nabla p$ in the force balance equation,
resulting in a non-symmetric matrix with consequences for the
construction of efficient solvers and preconditioners for the linear
system. 
This second issue also arises for any approximation of
equations~\eqref{eq:boussinesq-1}--\eqref{eq:boussinesq-2} that
includes a non-constant density in the mass balance equation, for
example the (truncated) anelastic liquid approximation
(T)ALA \cite{KLVLZTTK10}. Because of this universal importance, we
will discuss the difficulties that result from compressible models in
some detail in the following.

There are a number of possible avenues for linearisation of
\eqref{eq:boussinesq-2}. For example, one could instead use the equation
\begin{equation*}
  \nabla \cdot (\rho^\ast \ve u) = 0,
\end{equation*}
where $\rho^\ast$ is a known approximation of the density that is computed from the
previous time step's temperature and pressure, or from a temperature
and pressure that has been extrapolated from previous time steps to
the current time, and might be updated during a nonlinear iteration. 
Alternatively, in the case of the (T)ALA, 
$\rho^\ast = \rho^\ast(z)$ simply is a prescribed density profile 
that does not change over time. In all those cases, $\rho^\ast$ may still be spatially
variable, but it no longer depends on the quantities $\ve u,p$ that we
are currently solving for. While this
resolves the nonlinearity, the operator $-\nabla \cdot
(\rho^\ast \bullet)$ is not adjoint to the gradient operator acting on
the pressure in the force balance equation; direct discretizations of
this term therefore do not lead to a symmetric system matrix. 

In addition, the term is not computable in practice because the
product $\rho^\ast \mathbf u$ is not a finite element function (or
other polynomial) of
which we can compute derivatives during assembly. One way to make it
computable is to multiply out the
divergence. In order to make the equation look similar to the one we
have in the incompressible case, we also divide by the density. Two
choices that result from this are then to consider either%
\footnote{Both of these methods are also implemented in the widely
  used code CitcomS \cite{ZMTMG08}, though we are not aware of a
  systematic discussion of the two options, nor of comprehensive tests of
  their differences as we provide below.}
\begin{equation}
\label{eq:mass-implicit}
  \nabla \cdot \ve u + \frac{1}{\rho^\ast}\nabla \rho^\ast \cdot \ve u = 0,
\end{equation}
or
\begin{equation}
\label{eq:mass-explicit}
  \nabla \cdot \ve u = -\frac{1}{\rho^\ast}\nabla \rho^\ast \cdot \ve u^\ast,
\end{equation}
In the last equation, we have also frozen the velocity in the right hand side
term to a fixed
value obtained from previous time steps. If $\rho$ depends on the
pressure, either of these approaches then require a nonlinear
iteration to converge to the desired solution.

These two formulations are also not without difficulty. First, the
replacement $\nabla \cdot (\rho \ve u) = \nabla \rho \cdot \ve u +
\rho \nabla \cdot \ve u$ strictly only makes sense if the density is
continuous. If it is not, for example when taking into account phase
changes, then $\rho\ve u$ is a continuous function of which we can
take derivatives, whereas we cannot of its components $\rho$ and $\ve
u$.%
\footnote{This is, however, a theoretical consideration since the finite
element spaces we use will not allow us to represent discontinuous
velocities anyway.}
While the traditional approach to dealing with undesirable derivatives is to
multiply with test functions and integrate by parts, this is not
possible here because the pressure test functions with which this
equation is multiplied are only in $L_2$ and consequently not
sufficiently smooth to allow for integration by parts.

Second, there are also difficulties from the
perspective of finite element approximations when using a density 
$\rho=\rho(p,T)$ that depends on the primary variables pressure and
temperature (and possibly other variables such as the chemical
composition) in equation~\eqref{eq:mass-implicit} or \eqref{eq:mass-explicit}.
In this case,
$
  \nabla\rho(p,T)
  =
  \frac{\partial \rho}{\partial p}\nabla p
  +
  \frac{\partial \rho}{\partial T}\nabla T,
$
and likewise, for the finite element approximation (indicated by the index $h$),
$
  \nabla\rho(p_h,T_h)
  =
  \frac{\partial \rho}{\partial p}\nabla p_h
  +
  \frac{\partial \rho}{\partial T}\nabla T_h
$. On the other hand, the theory of the Stokes equations yields that
in general, the pressure is only a function in $L_2$, see for example \cite{guermond2004fem}. 
In
practice, this means that one does not usually get a better
approximation than $\|p-p_h\|_{L_2} = {\cal O}(h)$ for the finite
element approximation $p_h$ of the pressure, unless the solution happens to be
smooth. Indeed,
if the viscosity is discontinuous or has large gradients, one often
gets an even lower convergence order; for example, the SolCx test case
yields a convergence order $\|p-p_h\|_{L_2} = {\cal O}(h^{1/2})$ (see
\cite{KHB12}). This implies that, assuming we use a continuous
finite element space to approximate the pressure, we can at best hope that
$\nabla p_h$ converges to $\nabla p$ as $h\rightarrow 0$ in some
average sense, but that we can not expect this to happen with any
particular order; in other words, the best one might hope for is a
statement of the form $\|\nabla p-\nabla p_h\|_{L_2} = o(1)$,
but the approximation will likely be very poor and probably not
converge in a pointwise sense. (Indeed, we demonstrate this
experimentally in Section~\ref{sec:averaging}.) Pointwise convergence can obviously not
be expected at all if one uses discontinuous finite element spaces for
the approximation of the pressure.
Consequently, any practical scheme that
replaces $\nabla\rho(p_h,T_h)$ by terms that include $\nabla p_h$ will likely
yield a rather poorly approximated density gradient, resulting in
degradation in convergence of the velocity and temperature. We
therefore would like to avoid the occurrence of $\nabla p_h$ in our scheme.

To this end, we replace $\nabla p\approx \rho \mathbf g$. This is
motivated by the observation that for the hydrostatic pressure $p_s$
that dominates the total pressure in the Earth mantle, by definition
we have $\nabla p_s = \rho_\text{adi} \mathbf g$ with the adiabatic
reference density $\rho_\text{adi}$; indeed, in the (T)ALA
approximations, one chooses $\rho^\ast=\rho_\text{adi}$.
We can then approximate
$
  \nabla\rho(p,T)
  \approx
  \frac{\partial \rho}{\partial p}\rho \ve g
  +
  \frac{\partial \rho}{\partial T}\nabla T
$. Using this allows us to re-state the equations
above as
\begin{equation}
  \label{eq:div-approx-1-implicit}
  \nabla \cdot \ve u 
  + 
  \left(
    \frac{\partial \rho}{\partial p} \ve g
    +
  \frac{1}{\rho^\ast}
    \frac{\partial \rho}{\partial T}\nabla T^\ast
  \right) \cdot \ve u = 0,
\end{equation}
or
\begin{equation}
  \label{eq:div-approx-1-explicit}
  \nabla \cdot \ve u = 
  -
  \left(
  \frac{\partial \rho}{\partial p}\ve g
  +
  \frac{1}{\rho^\ast}
  \frac{\partial \rho}{\partial T}\nabla T^\ast
  \right) \cdot \ve u^\ast.
\end{equation}
In the following, we will call these two
options the \textit{implicit} and \textit{explicit approximation},
because they either include the velocity implicitly or explicitly in
the term that contains the gradient of the pressure.

Both of these approximations introduce
errors that depend on (i) how accurately $\rho^\ast=\rho(p^\ast,T^\ast)$ approximates
$\rho(p,T)$, which can be controlled by small time steps and accurate
extrapolations from previous time steps; and (ii) how good the
approximation for $\nabla p\approx \rho \mathbf g$ is, which is related to how small the
velocity is, and consequently how appropriate the choice of the
equations \eqref{eq:boussinesq-1}--\eqref{eq:boussinesq-2} was to
begin with. The more relevant question is therefore which of these
approximations one \textit{wants to use} for practical considerations.

To understand this, it is instructive to recall that discretizations
of the force balance equation~\eqref{eq:boussinesq-1} together with
the approximations \eqref{eq:div-approx-1-implicit}, 
and \eqref{eq:div-approx-1-explicit} lead to system
matrices with the following structure:
\begin{align*}
  \left(\begin{matrix}
    A & B^\T \\ B+C & 0
  \end{matrix}\right),
  \qquad
  \left(\begin{matrix}
    A & B^\T \\ B & 0
  \end{matrix}\right).
\end{align*}
Which one we choose has consequences for available choices of linear
solvers and preconditioners that are important since in most realistic
simulations, 70\% or more of the overall run time is spent in solving the
discretized velocity-pressure system. Furthermore, since we
linearise the equation we really want to solve, we will have to
iterate out the nonlinearity, and the two choices will
require different numbers of outer, nonlinear iterations. Predictably,
the choice that keeps the velocity entirely implicit,
\eqref{eq:div-approx-1-implicit}, and can therefore be expected to
converge more quickly in the nonlinear iteration, will also lead to
more difficult-to-solve linear systems due to the lack of
symmetry. Consequently, the choice between 
\eqref{eq:div-approx-1-implicit} and
\eqref{eq:div-approx-1-explicit} is not a priori clear.

\subsubsection{Correcting the right hand side}
\label{sec:compressibility-rhs}

When using the explicit approximation~\eqref{eq:div-approx-1-explicit}, 
we end up with an equation that is
rank deficient if the fluid flow is enclosed in a domain where the
normal component $b=\ve n \cdot \ve u$ of the fluid velocity is
prescribed on all parts of the boundary (a typical example being
either no-slip or tangential flow). In those cases, integrating over
the domain and using the divergence theorem yields
\begin{equation*}
  \int_{\partial\Omega} b
  = 
  -
  \int_\Omega
  \left(
  \frac{\partial \rho}{\partial p}\ve g
  +
  \frac{1}{\rho^\ast}
  \frac{\partial \rho}{\partial T}\nabla T^\ast
  \right) \cdot \ve u^\ast.
\end{equation*}
The left hand side of this equation is fixed and known based on the
given boundary conditions. On the other hand, the right hand side may
be whatever it is, based on our choice of approximations $T^\ast,\ve
u^\ast$ as well as the choice of quadrature formula and geometric
approximation of the domain. Thus, it may or may not equal the fixed value on the left,
and if it does not, then \eqref{eq:div-approx-1-explicit} will not allow for
a solution. On the other hand, it is clear that the difference between
the two sides will be small if $T^\ast,\ve u^\ast$ are well chosen and
if the assumptions that went into \eqref{eq:div-approx-1-explicit} are
valid. Thus, we can make the system solvable again by replacing
\eqref{eq:div-approx-1-explicit} by the equation
\begin{equation}
  \label{eq:div-approx-3-fixed}
  \nabla \cdot \ve u = 
  -
  \left(
  \frac{\partial \rho}{\partial p}\ve g
  +
  \frac{1}{\rho^\ast}
  \frac{\partial \rho}{\partial T}\nabla T^\ast
  \right) \cdot \ve u^\ast
  -
  \delta,
\end{equation}
where $\delta$ is chosen so that the invariant is always satisfied:
\begin{equation*}
  \delta
  = 
  -
  \frac 1{|\Omega|}
  \int_{\partial\Omega} b
  -
  \frac 1{|\Omega|}
  \int_\Omega
  \left(
  \frac{\partial \rho}{\partial p}\ve g
  +
  \frac{1}{\rho^\ast}
  \frac{\partial \rho}{\partial T}\nabla T^\ast
  \right) \cdot \ve u^\ast.
\end{equation*}
This correction $\delta$ is easily computed before assembling the
linear system that results from the linearisation of the Stokes equation, and
amounts to slightly correcting the compressibility everywhere to ensure global
mass conservation.

We note that in the case of an incompressible material, we have
$\frac{\partial \rho}{\partial p}=\frac{\partial \rho}{\partial T}=0$,
and mass conservation of course implies that the sum of influxes and
outfluxes has to balance, i.e., $\int_{\partial\Omega}
b=0$. Consequently, for incompressible materials, $\delta$ always evaluates to
zero; no correction is necessary in this case. 
(However, for inhomogeneous boundary conditions one has to be more careful, see \cite{heister_fluxpreserving}.)
Likewise, if the 
material is compressible but the setup of the problem has a part of
the boundary where only a normal stress of the fluid is prescribed,
then fluid velocity and pressure can adjust independently to allow any
right hand side to the mass conservation equation, and the correction
above is neither necessary nor desirable.

\subsubsection{Cost evaluation of the two formulations}
\label{sec:compressible-symmetry}

It is not a priori clear which of the two formulations,
\eqref{eq:div-approx-1-implicit} or \eqref{eq:div-approx-3-fixed}, is
preferable from a practical perspective: The first is ``more
implicit'' and consequently likely requires fewer nonlinear
iterations; the second yields a symmetric system matrix and
consequently likely requires fewer linear GMRES iterations because we
can formulate a better preconditioner.%
\footnote{A discussion of the preconditioner we use can be found in
  \cite{KHB12}. Specifically, for linear systems of the form
  $
  \left(\begin{matrix}
    A & B^\T \\ B+C & 0
  \end{matrix}\right),
  $
  we use the preconditioner proposed by Silvester and
  Wathen for the symmetric Stokes system (see \cite{SW94,ESW05} for a
  derivation):
  $
  P^{-1}=
  \left(\begin{matrix}
    \widetilde{A^{-1}} & \widetilde{A^{-1}}B^\T \widetilde{S^{-1}} \\ 0 & -\widetilde{S^{-1}}
  \end{matrix}\right),
  $
  where a tilde indicates an approximation and $S=B^TA^{-1}B$ is the
  Schur complement of the symmetric part. This preconditioner does not
  include the matrix $C$ and can therefore be expected to deteriorate
  if the compressibility in the implicit formulation becomes large.

  We have spent a significant amount of time testing preconditioners
  that include $C$ in some way, but have not been able to find ones
  that improve on the one shown above.
}
To resolve the question, we
have performed a number of numerical experiments.

\begin{figure}
 \centering
 \includegraphics[width=0.4\textwidth]{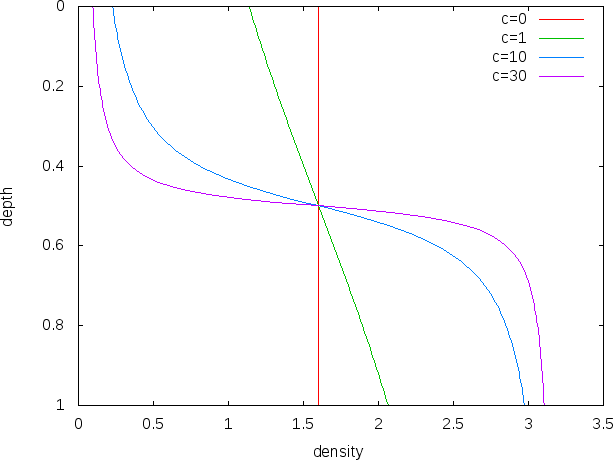}
 \caption{Density profiles used in the comparison between implicit and
   explicit formulations.}
 \label{fig:density-as-function-of-depth}
\end{figure}

Specifically, our test problem consists of a unit box, uses the
truncated anelastic liquid approximation (TALA), and a spatially
variable adiabatic density of the form
\[
 \bar{\rho}(z) = 1.6+\arctan\left(c(z-0.5)\right),
\]
where we will vary the coefficient that describes the deviation from a
constant density in the set $c \in \{0,1,10,30\}$ (see
Fig.~\ref{fig:density-as-function-of-depth}).
The density's derivative has a peak at $z=0.5$ with $\frac{d\rho}{dz}(0.5)=c$.
We use the non-dimensional Rayleigh number $\text{Ra}=10^4$ and dissipation
number $\text{Di}=0.1$,
and prescribe constant inflow at the top boundary,
$\mathbf{u}=(0,-1)$, free slip at left and right boundaries, and open
outflow at the bottom.

We show a comparison of the number of GMRES iterations in
Table~\ref{fig:implicit_ala_iter_table}. The numbers there show that
indeed a single implicit solve (using
\eqref{eq:div-approx-1-implicit}) is more expensive in terms of GMRES 
iterations than a single explicit solve (using
\eqref{eq:div-approx-3-fixed}) for all choices $c>0$ of the
compressibility parameter. In fact, iterations for a single solve of the explicit formulation
are independent of $c$.
On the other hand, the explicit formulation
requires a Picard iteration to iterate the nonlinearity, and the
number of linear GMRES iterations accumulated over these Picard
iterations is significantly larger than for the implicit formulation.

\begin{table}[ht]
  \centering
 \begin{tabular}{cr|rrrr}
  \multicolumn{6}{c}{implicit}  \\
  Mesh & DoFs & $c=0$ & $c=1$ & $c=10$ & $c=30$ \\ \hline
  $32\times 32$ & 9539 & 43 & 43 & 52 & 64 \\
  $64\times 64$ & 37507 & 46 & 48 & 62 & 75 \\
  $128\times 128$ & 148739 & 50 & 52 & 65 & 72 \\
  $256\times 256$ & 592387 & 56 & 62 & 79 & 114 \\
   &  &  &  &  &  \\
  \multicolumn{6}{c}{explicit} \\
  Mesh & DoFs & $c=0$ & $c=1$ & $c=10$ & $c=30$ \\ \hline
  $32\times 32$ & 9539 & 43 & 125 & 257 & 334 \\
  $64\times 64$ & 37507 & 46 & 133 & 289 & 388 \\
  $128\times 128$ & 148739 & 50 & 132 & 267 & 328 \\
  $256\times 256$ & 592387 & 56 & 140 & 303 & 407 \\
   &  &  &  &  &  \\
  \multicolumn{6}{c}{explicit (first nonlinear iteration)}  \\
  Mesh & DoFs & $c=0$ & $c=1$ & $c=10$ & $c=30$ \\ \hline
  $32\times 32$ & 9539 & 43 & 43 & 43 & 43 \\
  $64\times 64$ & 37507 & 46 & 46 & 46 & 46 \\
  $128\times 128$ & 148739 & 50 & 50 & 50 & 50 \\
  $256\times 256$ & 592387 & 56 & 56 & 56 & 56 \\
  \end{tabular}
  
%  \begin{tabular}{cr|rrr|rrr|rrr|rrr}
%   &  & \multicolumn{3}{c}{c=0} & \multicolumn{3}{c}{c=1} & \multicolumn{3}{c}{c=10} & \multicolumn{3}{c}{c=30}  \\
%  Mesh & DoFs & impl & expl & ex* & impl & expl & ex* & impl & expl & ex* & impl & expl & ex* \\ \hline
%  $32\times 32$ & 9539 & 43 & 43 & 43 & 43 & 125 & 43 & 52 & 257 & 43 & 64 & 334 & 43 \\
%  $64\times 64$ & 37507 & 46 & 46 & 46 & 48 & 133 & 46 & 62 & 289 & 46 & 75 & 388 & 46 \\
%  $128\times 128$ & 148739 & 50 & 50 & 50 & 52 & 132 & 50 & 65 & 267 & 50 & 72 & 328 & 50 \\
%  $256\times 256$ & 592387 & 56 & 56 & 56 & 62 & 140 & 56 & 79 & 303 & 56 & 114 & 407 & 56 \\
%  \end{tabular}

  \caption{Total linear GMRES solver iterations for implicit and explicit
    formulations. The explicit formulation requires an outer
    fixed-point iteration; the second set of numbers denotes the sum
    of linear iterations over all nonlinear iterations, whereas the
    third set of numbers denotes the number of linear
    iterations for the first nonlinear solver iteration.}
 \label{fig:implicit_ala_iter_table}
\end{table}

The result of these experiments is that for stationary computations,
the implicit formulation is both computationally cheaper and, likely,
more stable. On the other hand, for time dependent problems the
explicit formulation may be cheaper since one will already have a good
approximation for $\mathbf{u}^\ast$ and one may only need a single
nonlinear iteration.

\subsubsection{Benchmark for the compressible Stokes flow solver}
\label{sec:tan_gurnis}

We have verified our implementations of the compressible Stokes and
temperature formulations (Section~\ref{sec:compressible_stokes}) using a
number of benchmarks. In particular, we have reproduced the results from the
community benchmark described in \cite{KLVLZTTK10} (see Section~\ref{sec:compressible-2d-cartesian}). 
We have also reproduced the benchmark given in the Appendix
of \cite{TG07} and will describe our results in the following. This latter
benchmark consists of an analytical solution for an instantaneous compressible Stokes flow problem 
(with a given temperature). Using Fourier decomposition, the problem can be
reduced to a boundary value ordinary differential 
equation that can be solved numerically up to machine precision.

The test case in \cite{TG07} is defined in terms of the non-dimensional Rayleigh and dissipation
numbers,
\begin{align*}
\text{Di} = \frac{\alpha g L}{C_p}, \qquad\qquad
\text{Ra} = \frac{\alpha \Delta T \rho_0^2 g L^3 C_p}{\eta k},
\end{align*}
where $L$ a characteristic length scale, $\Delta T$ a characteristic
temperature difference, $\rho_0$ a reference density, and all other parameters
as introduced in Section~\ref{sec:formulation}. We then use the benchmark in
the form discussed in \cite{TG07}, but with equation~(B4) corrected to read
\begin{align*}
\frac{\text{Di}}{\text{Ra}} \underline{\sigma} : \underline{\varepsilon} = 
&\frac{\text{Di}}{\text{Ra}} \eta \left( 4 k^2 U_x^2 + \frac{10}{9} \beta^2 U_z^2 - 4 \beta k U_x U_z \right)  \cos^2(kx) \\
&+ \frac{\text{Di}}{\text{Ra}} \frac{1}{\eta} (\Sigma_{xz})^2 \sin^2(kx).
\end{align*}

We implement the benchmark in the setting of equations
\eqref{eq:boussinesq-1}--\eqref{eq:boussinesq-3} by fixing all of the above
material constants to $1$, except for $\alpha = \text{Di}$ and $\eta =
\text{Di}/\text{Ra}$. We then test both the Boussinesq approximation
(BA) and the truncated anelastic liquid approximation (TALA),
and compute the $L_2$ error of the velocity, and errors of the integrals of
shear ($W = \tau(\ve u) : \varepsilon(\ve u)$) and adiabatic heating 
($\phi = \alpha \rho T (\ve u \cdot \ve g)$). 
The problem is instantaneous, so we perform a nonlinear iteration
with the explicit formulation of the compressibility for a single timestep.
Alternatively, one can use the implicit formulation and perform a single
Stokes solve, which gives very similar results.

The results are shown in
Table~\ref{fig:tangurnis} and %Figure~\ref{fig:tangurnis1} and
show optimal third order convergence for the $L_2$ error of the velocity. 
Both heating terms show less regular, but equally fast convergence to the 
exact values, with the total shear heating converging at an even higher order 
than the velocity.

\begin{table}
 \centering

 \begin{tabular}{r|r|r|rr}
  1/h & $\|\mathbf u-\mathbf u^*\|_0$ & rate & $|W-W^*|$ & $|\phi-\phi^*|$ \\\hline \hline
  \multicolumn{5}{c}{Boussinesq approximation (BA)} \\ \hline
  8 & 9.0721e-06 & - &  &  \\
  16 & 1.1103e-06 & 3.03 &  &  \\
  32 & 1.3806e-07 & 3.01 &  &  \\
  64 & 1.7242e-08 & 3.00 &  &  \\
  %128 & 2.2451e-09 & 2.94 &  &  \\
  \hline
  \multicolumn{5}{c}{Truncated anelastic liquid approximation (TALA), $a=0$} \\ \hline
  8 & 1.2109e-05 & - & 4.5439e-07 & 2.2179e-07 \\
  16 & 1.4840e-06 & 3.03 & 2.9067e-08 & 1.4130e-08 \\
  32 & 1.8459e-07 & 3.01 & 1.6974e-09 & 5.5979e-10 \\
  64 & 2.3056e-08 & 3.00 & 6.2599e-11 & 2.9021e-10 \\
  %128 & 2.9842e-09 & 2.95 & 1.8260e-10 & 3.5021e-10 \\
  \hline
  \multicolumn{5}{c}{Truncated anelastic liquid approximation (TALA), $a=2$} \\ \hline
  8 & 8.7973e-06 & - & 2.1267e-07 & 1.4399e-07 \\
  16 & 1.1207e-06 & 2.97 & 1.3707e-08 & 9.3239e-09 \\
  32 & 1.4078e-07 & 2.99 & 8.0734e-10 & 4.9389e-10 \\
  64 & 1.7638e-08 & 3.00 & 2.6633e-12 & 6.6108e-11 \\
  %128 & 2.2702e-09 & 2.96 & 5.2663e-11 & 1.0611e-10 \\
  \end{tabular}
  
 \caption{Convergence of velocity and heating terms for the benchmark problem
   defined in \cite{TG07}. The exact values $\mathbf u^\ast,W^\ast,\phi^\ast$
   are known from the exact solution of the problem.}
 \label{fig:tangurnis}
\end{table}

% \begin{figure}
%  \centering
%  \includegraphics[width=0.4\textwidth]{tangurnis/tangurnis.pdf}
%  \caption{Convergence of the $L_2$ error of $x$ component of the velocity for
%    the benchmark problem defined in \cite{TG07}.}
%  \label{fig:tangurnis1}
% \end{figure}

\subsubsection{Benchmark for 2d Cartesian compressible convection}
\label{sec:compressible-2d-cartesian}

In order to verify that our approaches to solving compressible
problems also work for more complex applications, we also evaluate the
correctness and accuracy of the re-formulations of the equations
introduced in Section~\ref{sec:compressible_stokes} using the
community benchmark defined in \cite{KLVLZTTK10}.
The model domain for this benchmark is a 2-D square box cooled from the top and heated from the bottom. This setup corresponds to the benchmark given in \cite{BBC89}, except that the material is no longer assumed to be incompressible and instead different approximations for the compressible mass conservation equation are tested. All material properties are approximated as constants, with the exception of the density, which varies around a reference state
\begin{equation}
\bar{\rho} = \rho_0 \exp \left( z \frac{\text{Di}}{\gamma} \right).
\end{equation}
A constant temperature is prescribed at the top ($z=0$) and bottom ($z=L$) of the domain, with
\begin{align*}
T_\text{top} &= \frac{T_\text{surf}}{\Delta T}, \qquad &
T_\text{bot} &= \frac{T_\text{surf} + \Delta T}{\Delta T}
\end{align*}
and no flux conditions at the side walls. This temperature increase across the model domain includes both the contribution of the adiabatic temperature profile,
\begin{equation}
\bar{T} = T_\text{top} \; e^{z \, \text{Di}},
\end{equation}
and the nonadiabatic temperature variations across the boundary layers. 
The initial temperature is a linear profile that
matches these boundary conditions, plus a small perturbation:
\begin{equation*}
T_{t=0} = \frac{z}{L} + 0.01 \cos\left(\frac{\pi x}{L}\right) \sin\left(\frac{\pi z}{L}\right) + T_\text{top}.
\end{equation*}
We then let the model evolve until steady state is reached. 

Analogous to the procedure described in Section~\ref{sec:tan_gurnis}, we reproduce the non-dimensional formulation of the benchmark 
by setting all material constants to 1, except for $\alpha = \text{Di}$ and $\eta = \text{Di}/\text{Ra}$. 
In the different benchmark cases, $\text{Di}$ is varied between $0.25$ and $1$, and $\text{Ra}$ is chosen as $10^4$ and $10^5$. 
All parameters are given in Table~\ref{table:king}.

\begin{table}
 %\centering
 \resizebox{\columnwidth}{!}{%
 \begin{tabular}{lll}
  \hline
  & Expression  & Value \\ \hline
  $\Delta T$ & temperature change across the domain & 3000\,K \\
  $T_\text{surf}$ & surface temperature & 273\,K \\
  $\gamma$ & Grueneisen parameter& 1 \\
  $L$ & width and height of the box & 1\,m \\
  $g$ & gravitational acceleration in negative z direction & 1\,m\,s\textsuperscript{-2} \\
  $\alpha$ & thermal expansivity & Di \\
  $c_p$ & specific heat & 1 J\,kg\textsuperscript{-1}\,K\textsuperscript{-1}\\
  $\rho_0$ & surface density & 1 kg\,m\textsuperscript{-3}\\
  $\eta$ & viscosity & Di/Ra \\
  $k$ & thermal conductivity & 1 W\,m\textsuperscript{-1}\,K\textsuperscript{-1}\\
  \hline
  \end{tabular}
   }
 \caption{Parameters for the benchmark defined in \cite{KLVLZTTK10}.}
 \label{table:king}
\end{table}

We have tested both the anelastic liquid approximation (ALA) and the
truncated anelastic liquid approximation (TALA) using our reference
implementation of our algorithms in the \aspect{} code \cite{KHB12}.
Because we are only interested in the steady-state limit, rather than
accurate intermediate values, we report results for the explicit formulation
\eqref{eq:div-approx-1-explicit} with the modification in
\eqref{eq:div-approx-3-fixed}, without actually iterating out the
nonlinearity in every time step. (However, we have also verified that the implicit
formulation, \eqref{eq:div-approx-1-implicit}, yields essentially the
same results.)
Table~\ref{table:king_results_short} provides an excerpt of
results for the ALA, with full results for both ALA and TALA given in
Tables~\ref{fig:king_full_results_ala} and
\ref{fig:king_full_results_tala}. Specifically, we compare the Nusselt
number $\text{Nu}$, root mean square velocity $V_\text{rms}$, average
temperature $\left< T \right>$, the total shear heating $\phi$ and
adiabatic heating $W$ to the results given in \cite{KLVLZTTK10}. As
can be seen from the table, there is excellent agreement between our
results and those previously reported. In other words, the
re-formulations in Section~\ref{sec:compressible_stokes} do not only
allow us to \textit{efficiently} solve compressible problems, but also
\textit{accurately}.

\begin{table*}
\centering
\begin{tabular}{llllllll}
  \hline
  Di & Ra &  & Nu & Vrms & $\left<T\right>$ & $\phi$ & W \\
  \hline \rowcolor{Gray} \rule{0pt}{2.2ex}
  0.25 & $10^4$ & \aspect{} & 4.4145 & 39.9568 & 0.5149 & 0.8496 & 0.849 \\
   &  & King UM & 4.406 & 39.952 & 0.515 & 0.847 & 0.849 \\
   &  & King VT & 4.4144 & 40.0951 & 0.5146 & 0.849 & 0.849 \\
   &  & King CU & 4.41 & 40 & 0.5148 & 0.8494 & 0.8501 \\
%   0.5 & $10^4$ & \aspect{} & 3.822 & 35.9394 & 0.5224 & 1.3836 & 1.3819 \\
%    &  & King UM & 3.812 & 35.936 & 0.522 & 1.381 & 1.381 \\
%    &  & King VT & 3.8218 & 36.0425 & 0.5214 & 1.3812 & 1.3812 \\
%    &  & King CU & 3.82 & 35.9 & 0.5217 & 1.3818 & 1.383 \\
  \hline \rowcolor{Gray} \rule{0pt}{2.2ex}
  1 & $10^4$ & \aspect{} & 2.446 & 24.6809 & 0.5114 & 1.3427 & 1.354 \\
   &  & King UM & 2.438 & 24.663 & 0.512 & 1.343 & 1.349 \\
   &  & King VT & 2.4716 & 25.0157 & 0.51 & 1.3622 & 1.3621 \\
   &  & King CU & 2.47 & 24.9 & 0.5103 & 1.3627 & 1.3638 \\
  \hline \rowcolor{Gray} \rule{0pt}{2.2ex}
  0.25 & $10^5$ & \aspect{} & 9.2334 & 178.0751 & 0.5322 & 2.0525 & 2.0517 \\
   &  & King UM & 9.196 & 178.229 & 0.532 & 2.041 & 2.051 \\
   &  & King VT & 9.2428 & 179.7523 & 0.5318 & 2.0518 & 2.0519 \\
   &  & King CU & 9.21 & 178.2 & 0.5319 & 2.0503 & 2.054 \\
%   0.5 & $10^5$ & \aspect{} & 7.5674 & 155.1196 & 0.548 & 3.2368 & 3.2351 \\
%    &  & King UM & 7.532 & 155.304 & 0.548 & 3.221 & 3.233 \\
%    &  & King VT & 7.5719 & 156.5589 & 0.5472 & 3.2344 & 3.2346 \\
%    &  & King CU & 7.55 & 155.1 & 0.5472 & 3.233 & 3.2392 \\
  \hline \rowcolor{Gray} \rule{0pt}{2.2ex}
  1 & $10^5$ & \aspect{} & 3.8699 & 84.3678 & 0.5298 & 2.7519 & 2.7692 \\
   &  & King UM & 3.857 & 84.587 & 0.53 & 2.742 & 2.765 \\
   &  & King VT & 3.878 & 85.5803 & 0.5294 & 2.761 & 2.7614 \\
   &  & King CU & 3.88 & 84.6 & 0.5294 & 2.7652 & 2.7742 \\ \hline 
  \end{tabular}
  \caption{Excerpt of benchmark results for ALA as defined in
    \cite{KLVLZTTK10}. The \aspect{} computations are highlighted in
    gray and were obtained using extrapolation from a 1/128
    mesh. Acronyms for the different codes used in \cite{KLVLZTTK10}
    are UM -- University of Michigan (Sepran); VT -- Virginia Tech
    (ConMan); CU -- University of Colorado at Boulder (Citcom). See
    the appendix for the full results.}
 \label{table:king_results_short}
\end{table*}

\subsection{Averaging of material properties}
\label{sec:averaging}

Geophysical models are often characterized by abrupt and large jumps in material
properties, in particular in the viscosity. An example is a subducting, cold
slab surrounded by the hot mantle: Here, the strong
temperature-dependence of the viscosity will lead to a sudden jump in the
viscosity between mantle and slab. Another example are phase transitions, 
where the density and viscosity of rocks change abruptly between the stability
field of different minerals. The length scale over which this happens
will be a few or a few tens of kilometres. Such length scales cannot be
adequately resolved in three-dimensional computations with typical meshes for
global computations. In other words, the viscosity field is, for all
practical purposes, discontinuous, with jumps of possibly several orders of
magnitude from quadrature point to quadrature point.

Having large viscosity variations in models poses a variety of problems to
numerical computations. First, they lead to very long compute
times because solvers and/or preconditioners break down (see \cite{rudi2015extreme} for a proposed
preconditioner for large viscosity variations). This may be
acceptable if it would at least lead to accurate solutions, but large viscosity
variations also lead to large pressure gradients, and this in turn leads to over-
and undershoots in the numerical approximation of the gradient. We will 
demonstrate both of these issues experimentally in Section~\ref{sec:averaging_accuracy} and \ref{sec:averaging_speed} below.

One can mitigate some of these effects by averaging material
properties in some form on each cell (see, for example,
\cite{Bab08,Deu08,DMGT11,Thi15,TMK14}). At the same time, replacing
the \textit{correct} viscosity at each quadrature point by an
\textit{averaged} one implies solving a different problem, and one
would expect this to affect the accuracy of the solution. In cases
where the viscosity (and consequently the solution) is smooth,
averaging could be assumed to be harmful to the overall accuracy. On
the other hand, if the solution has essentially discontinuous
gradients and kinks in the velocity field, then at least at these
locations we cannot expect a particularly high convergence order
anyway, and the averaging will likely not hurt very much either. This
section
therefore explores these issues and shows numerical results.

\subsubsection{Implementation}

In implementations, averaging first
evaluates the material model at every quadrature point of a cell,
given the temperature, pressure, strain rate, and other quantities at
these points, and then either (i)~uses these values as is in the
assembly of contributions to the system matrix and right hand side, or
(ii) replaces the values by their arithmetic average $\bar x = \frac 1N
\sum_{i=1}^N x_i$, harmonic average $\bar x = \left(\frac 1N
\sum_{i=1}^N \frac{1}{x_i}\right)^{-1}$, geometric average $\bar x =
\left(\prod_{i=1}^N \frac{1}{x_i}\right)^{-1/N}$, or largest value
over all quadrature points on this cell.  Alternatively, one may
project the values from the quadrature points to a bi- (in 2d) or
trilinear (in 3d) $Q_1$ finite element space on every cell, and then
evaluate this finite element representation again at the quadrature
points; in this case, one may also limit the computed values at quadrature
points by the minimum and maximum value of the coefficient before
averaging. These operations are applied to all quantities that the
material model computes, 
i.e., in particular, the viscosity, the density, the compressibility, and the
various thermal and thermodynamic properties. 

A priori, we know of little guidance from the literature on the
analysis of numerical discretizations of partial differential equations regarding
the question which of these averaging options is best. Indeed, it is
also not quite clear what the appropriate metric would be to determine
``best'' -- for example, one could consider various norms of the errors, run time
of solvers, or other metrics. Consequently, in the following sections we will consider a simple
test case and evaluate the options above with regard to discretization
error and the time necessary to solve the linear system associated
with each.

\subsubsection{Influence of averaging on numerical accuracy}
\label{sec:averaging_accuracy}

We experimentally evaluate the question which of the introduced averaging operations 
may in fact be best by considering the
``sinker'' benchmark. This benchmark is defined by a high-viscosity,
heavy disk at the center of a two-dimensional box. Both density and
viscosity are therefore discontinuous along the interface of the
disk, and in particular not aligned with the mesh. We use $\rho=1,
\eta=1$ outside the disk, and $\rho=10,\eta=10^6$ inside the disk
to simulate a realistic viscosity contrast; the contrast in the
density is immaterial as it is only a (global) scaling factor for the
solution.

\begin{figure}
  \centering
  \begin{tabular}{cccccc}
    \includegraphics[width=0.14\textwidth]{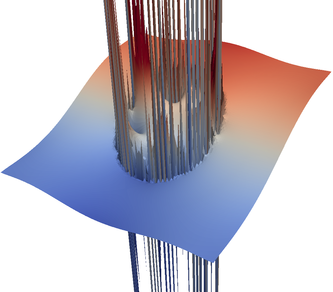}
    &
    \includegraphics[width=0.14\textwidth]{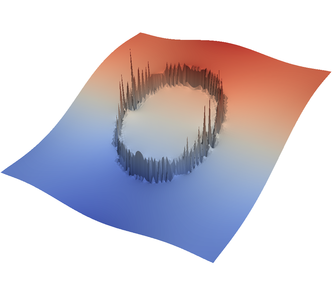}
    &
    \includegraphics[width=0.14\textwidth]{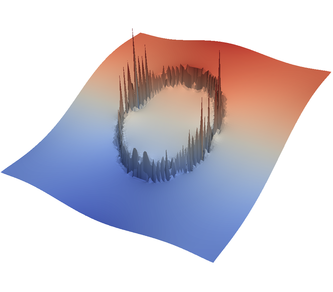}
%    &
%    \includegraphics[width=0.14\textwidth]{graphics/averaging/q2q1/sinker-7-geometric.png}
%    &
%    \includegraphics[width=0.14\textwidth]{graphics/averaging/q2q1/sinker-7-largest.png}
%    &
%    \includegraphics[width=0.14\textwidth]{graphics/averaging/q2q1/sinker-7-project.png}
    \\
    $[-45.2,45.2]$
    &
    $[-2.67,2.67]$
    &
    $[-3.58,3.58]$
%    &
%    $[-3.57,3.57]$
%    &
%    $[-1.80,1.80]$
%    &
%    $[-2.77,2.77]$
    \\
    \\
    \includegraphics[width=0.14\textwidth]{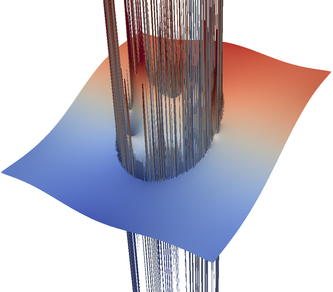}
    &
    \includegraphics[width=0.14\textwidth]{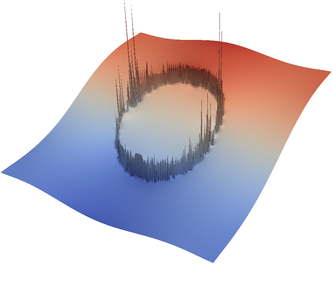}
    &
    \includegraphics[width=0.14\textwidth]{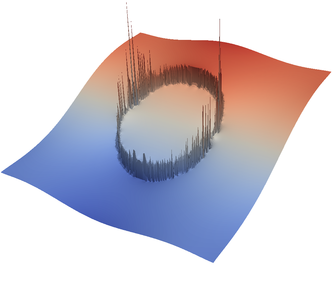}
%    &
%    \includegraphics[width=0.14\textwidth]{graphics/averaging/q2q1/sinker-8-geometric.png}
%    &
%    \includegraphics[width=0.14\textwidth]{graphics/averaging/q2q1/sinker-8-largest.png}
%    &
%    \includegraphics[width=0.14\textwidth]{graphics/averaging/q2q1/sinker-8-project.png}
    \\
    $[-44.5,44.5]$
    &
    $[-5.18,5.18]$
    &
    $[-5.09,5.09]$
%    &
%    $[-5.18,5.18]$
%    &
%    $[-5.20,5.20]$
%    &
%    $[-7.99,7.99]$
  \end{tabular}
  \caption{\it Visualization of the pressure field for the ``sinker''
    problem. Left to right: 
    No averaging, 
    arithmetic averaging, 
    harmonic averaging.
    %geometric averaging, 
    %pick largest.
    %project to $Q_1$.
    Top: On a mesh with $128\times 128$ cells. Bottom: On a
    mesh with $256\times 256$ cells. The minimal and maximal pressure
    values are indicated below every picture. This range is symmetric because
    we enforce that the average of the pressure equals zero. The color scale
    is adjusted to only show values between $p=-3$ and
    $p=3$. (Geometric averaging, choosing the largest value on each
    cell, and projecting the coefficient to a $Q_1$ space yields
    similar pictures, with pressure ranges
    $[-3.57,3.57]$, $[-1.80,1.80]$, and $[-3.58,3.58]$ for the coarser
    of the two meshes, and $[-5.18,5.18]$, $[-5.20,5.20]$, and
    $[-7.99,7.99]$ for the finer one.)}
  \label{fig:sinker-with-averaging-pressure}
\end{figure}

For three of the averaging options introduced above, and for different levels of mesh
refinement, Fig.~\ref{fig:sinker-with-averaging-pressure} shows
pressure plots that illustrate the problem with oscillations of the discrete
pressure, without and with averaging. The important part of these
plots is not that the solution looks 
discontinuous -- in fact, the exact solution is discontinuous at the edge of the
circle --
but the spikes that go far above and below the ``cliff'' in the pressure along 
the edge of the circle. Without averaging, these spikes are far
larger than the actual jump height. Importantly, the spikes also do not disappear 
under mesh refinement nor averaging; in other words, the discrete
pressure does not converge in the $L_\infty$ norm to the exact
pressure. (Further investigations also show that the maximal and
minimal pressures continue to grow with mesh refinement, although
slowly, with or without averaging.) On the other hand, the pressure
spikes become far less pronounced with averaging.

\begin{figure}
  \centering
    \includegraphics[width=0.23\textwidth]{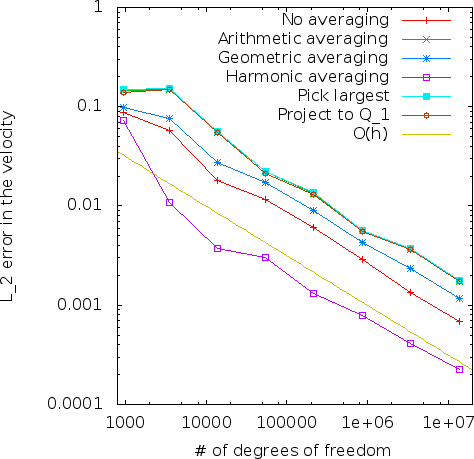}
    \hfill
    \includegraphics[width=0.23\textwidth]{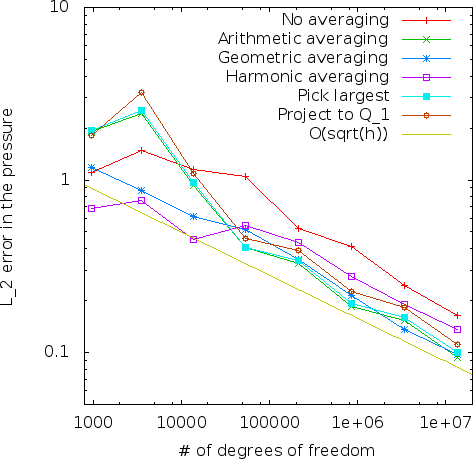}
  \caption{\it $L_2$ errors in velocity (left) and pressure (right)
    with a variety of averaging schemes as a function of the number of
    unknowns in the discretization. The figures shown here use the
    usual Taylor-Hood $Q_2^d\times Q_1$ element.}
  \label{fig:sinker-with-averaging-errors}
\end{figure}

The results shown in the figure do not allow to draw
definitive conclusions as to which averaging approach is the
best. This is in line with previous discussions of this
question, for example in \cite{Bab08,Deu08,DMGT11,TMK14}). On the
other hand, we can investigate this by evaluating the error in the
solution for the closely related ``Pure shear/Inclusion'' benchmark
(see \cite{DMGT11}) for which we know the exact solution.
To this end, Fig.~\ref{fig:sinker-with-averaging-errors} shows the
$L_2$ errors in velocity and pressure for a variety of averaging
options and as meshes are refined. The figures clearly show that all averaging schemes improve
the pressure approximation, though some deteriorate the velocity
approximation. In light of Figures
\ref{fig:sinker-with-averaging-pressure} and
\ref{fig:sinker-with-averaging-errors}, using harmonic averaging
appears to be a reasonable compromise. This is again in agreement with
previous statements in the literature.

\begin{figure}
  \centering
  \begin{tabular}{cccccc}
    \includegraphics[width=0.14\textwidth]{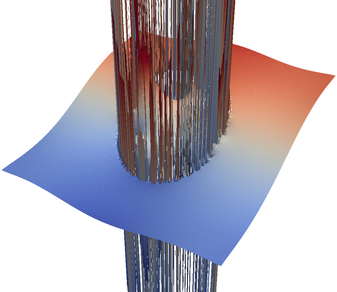}
    &
    \includegraphics[width=0.14\textwidth]{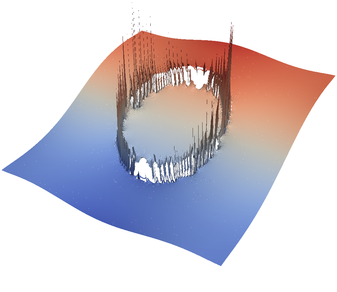}
    &
    \includegraphics[width=0.14\textwidth]{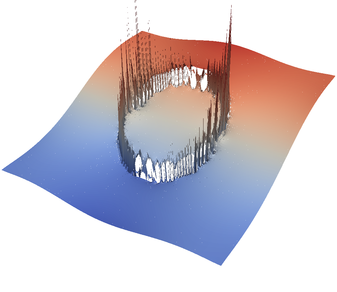}
%    &
%    \includegraphics[width=0.14\textwidth]{graphics/averaging/q2q1plus/sinker-7-geometric.png}
%    &
%    \includegraphics[width=0.14\textwidth]{graphics/averaging/q2q1plus/sinker-7-pick-largest.png}
%    &
%    \includegraphics[width=0.14\textwidth]{graphics/averaging/q2q1plus/sinker-7-project-to-Q1.png}
    \\
    \includegraphics[width=0.14\textwidth]{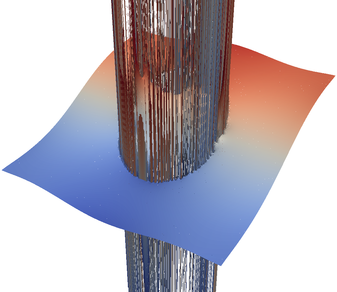}
    &
    \includegraphics[width=0.14\textwidth]{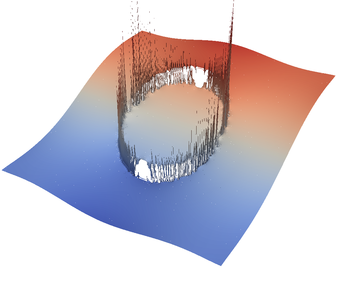}
    &
    \includegraphics[width=0.14\textwidth]{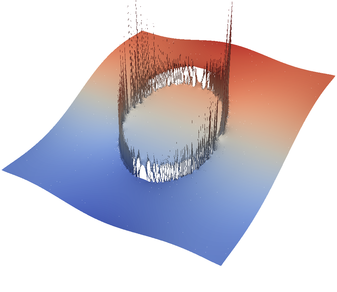}
%    &
%    \includegraphics[width=0.14\textwidth]{graphics/averaging/q2q1plus/sinker-8-geometric.png}
%    &
%    \includegraphics[width=0.14\textwidth]{graphics/averaging/q2q1plus/sinker-8-pick-largest.png}
%    &
%    \includegraphics[width=0.14\textwidth]{graphics/averaging/q2q1plus/sinker-8-project-to-Q1.png}
  \end{tabular}
  \caption{\it Visualization of the pressure field for the ``sinker''
    problem. Like Fig.~\ref{fig:sinker-with-averaging-pressure} but using the
    Stokes element with discontinuous pressures.}
  \label{fig:sinker-with-averaging-pressure-q2q1iso}
\end{figure}

\begin{figure}
  \centering
    \includegraphics[width=0.23\textwidth]{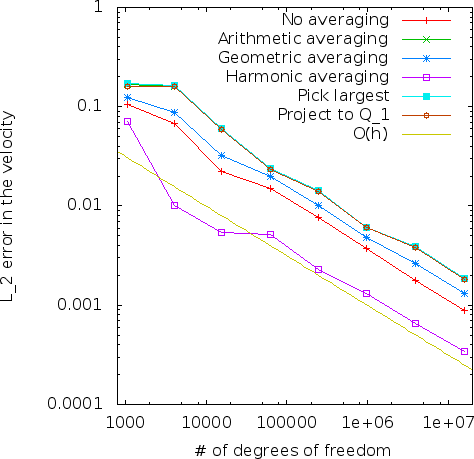}
    \hfill
    \includegraphics[width=0.23\textwidth]{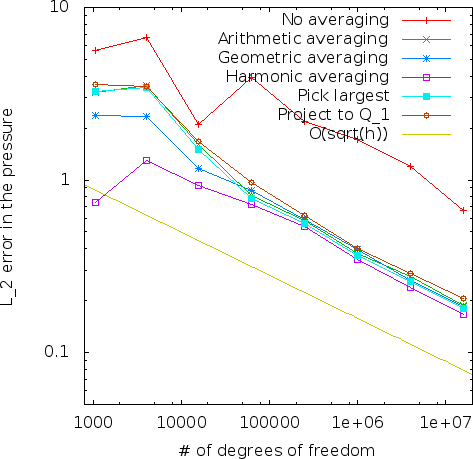}
  \caption{\it $L_2$ errors in velocity (left) and pressure (right)
    with a variety of averaging schemes as a function of the number of
    unknowns in the discretization. Compared to
    Fig.~\ref{fig:sinker-with-averaging-errors}, the graphs shown here
    use a Stokes element with a discontinuous pressure $Q_2^d\times P_{-1}$.}
  \label{fig:sinker-with-averaging-errors-q2q1iso}
\end{figure}

One may follow the problem with discontinuous pressures in a different
direction and suggest that the pressure could be better approximated
by using a discontinuous pressure space. This is in fact possible for
the Stokes equations, by choosing a discontinuous $P_k$ pressure space
instead of the common continuous $Q_k$ space of the Taylor-Hood pair,
without losing the inf-sup stability of the
discrete problem \cite{KHB12}.
Disappointingly, however, this makes no real difference: the pressure
oscillations are no better (in fact, they are worse) than for the standard Stokes
element (Fig.~\ref{fig:sinker-with-averaging-pressure-q2q1iso}) and
the $L_2$ errors are generally worse for both velocity and pressure
(Fig.~\ref{fig:sinker-with-averaging-errors-q2q1iso}).

\subsubsection{Influence of averaging on solver speed}
\label{sec:averaging_speed}

A very pleasant side effect of averaging is that solutions are not only
better behaved, but are also cheaper to compute.
For example, the total run time for the sinker testcase
of the previous section (see
Fig.~\ref{fig:sinker-with-averaging-pressure}), using a $256\times 256$ mesh and the
Taylor-Hood element, is reduced from 5870s without averaging to 240s for harmonic
averaging -- a speed-up of a factor of around 25!
% sinker/time_q2q1.prm, mpirun -n 1, release mode -> time_q2q1.log

Such improvements carry over to more complex and realistic models. For
example, in a simulation with large viscosity heterogeneities
using approximately 17 million unknowns and run on 64 processors, the
wall-clock run time is reduced from 145 hours to 17
hours and the computed solutions do not differ in any significant way.

We attempt to quantify this effect in
Table~\ref{tab:sinker-with-averaging-iteration-counts} by looking at
the number of outer GMRES iterations necessary to solve the variable
viscosity Stokes system. We use a preconditioner
equations that involves an inner solver with an
algebraic multigrid preconditioner for the
elliptic top-left block of the matrix (corresponding to the ``expensive''
option discussed in \cite{KHB12}). Using this scheme, the number
of GMRES iterations rises steeply with mesh refinement without
averaging, see Table~\ref{tab:sinker-with-averaging-iteration-counts}. On the
other hand, with (any kind of) averaging, the number of iterations
remains much lower. The effect is even more dramatic when using the
discontinuous pressure element mentioned in the previous section:
there, without averaging, the number of iterations grows from 389
on a $16\times 16$ mesh to 1174 on a $256\times 256$ mesh, while the number of
iterations with averaging are very similar to those shown in
Table~\ref{tab:sinker-with-averaging-iteration-counts}. 
% sinker/iter_dgp1.prm, mpirun -n 1, release mode, uses expensive stokes solver,
% (cheap 1000 gives 1000+396). see iter_dgp1.log

We can also quantify how many fewer outer GMRES iterations one needs
with averaging for the complex model mentioned above:
There, the number of iterations is reduced from 169 to 77.

However, the number of outer GMRES iterations is only part
of the problem. Depending on the choice of preconditioner for
the Stokes system, one has to also iteratively invert the elliptic
top-left block of the Stokes matrix, and/or a pressure mass
matrix. These ``inner'' solves also become vastly cheaper with
averaging, requiring 2 to 5 times fewer Conjugate Gradient
iterations than without averaging per preconditioner
application. Together with the reduction in outer iterations, overall
run time for the Stokes solver is reduced by the factors discussed at
the beginning of the subsection.

\begin{table}
  \center
  \begin{tabular}{|c|ccc|}
    \hline
    Mesh size
    & No averaging 
    & Arithmetic 
    & Harmonic 
    \\
    & 
    & averaging 
    & averaging 

    \\ \hline
  $16\times 16$ & 60 & 25 & 20 \\
  $32\times 32$ & 89 & 24 & 22 \\
  $64\times 64$ & 129 & 24 & 24 \\
  $128\times 128$ & 138 & 26 & 24 \\
  $256\times 256$ & 277 & 25 & 25 \\ \hline
  % see sinker/iters.prm and iters.log, expensive only
  \end{tabular}
  \caption{\it Number of outer GMRES iterations to solve the Stokes equations
  with continuous pressure
  on a sequence of globally refined meshes and for different
  material averaging operations. Geometric averaging, picking the
  largest viscosity value on each cell, and projecting the viscosity
  field to a piecewise $Q_1$ space yields very similar numbers as the
  other two averaging options.
  For an interpretation of the data
  see the main text.}
  \label{tab:sinker-with-averaging-iteration-counts}
\end{table}

\subsection{Latent heat}
\label{sec:latentheat}

When incorporating phase transitions into realistic mantle convection models we are not only faced with abrupt 
changes of material properties across these transitions as discussed in Section~\ref{sec:averaging}, 
but also with a relatively sudden change in internal energy of the material. This means that latent heat is 
consumed or released over a sharp interface as material crosses a particular phase boundary. 
In the energy balance \eqref{eq:boussinesq-3}, this is expressed as a heating term describing 
the changes of the entropy $S$ in terms of its material derivative. As the entropy of a given 
material depends only on temperature and pressure (assuming a constant
chemical composition), we can rewrite the corresponding heating term in \eqref{eq:boussinesq-3} as
\begin{align*}
\rho T \frac{\mathrm{D} S}{\mathrm{D} t} 
&=
\rho T \left(\frac{\partial S}{\partial T} \frac{\mathrm{D} T}{\mathrm{D} t} 
+ \frac{\partial S}{\partial p} \frac{\mathrm{D} p}{\mathrm{D} t} \right)
\\
&=
\rho T \left(\frac{\partial S}{\partial T} \left(\frac{\partial T}{\partial t} + \mathbf u\cdot\nabla T\right)
+ \frac{\partial S}{\partial p} \left(\frac{\partial p}{\partial t} + \mathbf u\cdot\nabla p\right) \right)
\end{align*}
Together with the approximation that the fluid is anelastic (see
Section~\ref{sec:formulation}) -- that is, assuming
$\frac{\partial p}{\partial t}=0$ -- and when moving all advection terms involving the temperature to the 
left-hand side, the energy balance \eqref{eq:boussinesq-3} can be rewritten in
the following form:
\begin{equation}
  \label{eq:temp-latent-heat}
  \begin{split}
  \left(\rho C_p - \rho T \frac{\partial S}{\partial T}\right) 
  \left(\frac{\partial T}{\partial t} + \mathbf u\cdot\nabla
  T\right) - \nabla\cdot k\nabla T \qquad
  \\
  \qquad
  =
  \rho H + \tau(\ve u) : \varepsilon(\ve u)
  +\left( \alpha + \rho \frac{\partial S}{\partial p} \right) T \left( \mathbf u \cdot \nabla p \right).
  \end{split}
\end{equation}

\subsubsection{Implementation}
Different approaches for how to implement this equation have been suggested in 
the literature:
\begin{enumerate}
\item One may describe a number of prominent phase transitions using the Clapeyron slope $\gamma$, density change $\Delta\rho$
      and an analytic phase function $X$, such as a hyperbolic tangent, that describes the stability field
      of each phase and varies between 0 and 1, 
      \[
      \frac{\partial S}{\partial T} = \Delta S \frac{\partial X}{\partial T}
      = \gamma \frac{\Delta\rho}{\rho^2} \frac{\partial X}{\partial T},
      \]
      see for example \cite{CY85}.
\item One may use a thermodynamic calculation package, such as Perple\_X \cite{C09} or BurnMan \cite{cottaar2014burnman}
      to compute $p$-$T$ tables of material properties, including the enthalpy $H$ (or its pressure and temperature derivatives),
      which describes the energy changes associated with phase
      transitions. Between data points of these tables, one may then
      interpolate continuously (yielding a smoothed out approximation of the
      true $p$-$T$ diagram) and compute derivatives
      $\partial S / \partial T$ and $\partial S / \partial p$ based on this interpolation.
\item One may use a modified version of (ii) that involves using the pressure and temperature derivatives 
      of the enthalpy to compute an ``effective'' thermal expansivity 
      \[ \alpha_\text{eff}= \frac{1}{T} \left[ 1 - \rho \left(\frac{\partial H}{\partial p} \right)_T \right] \]
      and specific heat 
      \[C_{p,\text{eff}} = \left(\frac{\partial H}{\partial T} \right)_p, \] 
      respectively, which are then used in the energy conservation equation in place of the original 
      quantities and account for latent heat effects (see for example \cite{NTDC09}). 
\end{enumerate}

All of these methods have in common that they introduce relatively narrow 
regions where latent heat is consumed or released. 
Even though phase changes generally occur over a range of pressures and 
temperatures, and are also not instantaneous, their width is often below 
the grid resolution of geodynamic computations. Hence, strategies have 
to be designed for smoothing out sharp transitions so that they can be 
treated numerically, but still yield a high accuracy.
In addition, narrow zones of latent heat release lead to strong temperature gradients
with consequent difficulties for numerical schemes that have to be addressed
by stabilization as discussed in \cite{KHB12}.

\subsubsection{Numerical results}
We numerically evaluate the reformulation of latent heat processes in
\eqref{eq:temp-latent-heat} by using the benchmark described in
\cite[part 1, p. 194]{STO01}.  It provides an analytical solution for
the latent heat that is released or consumed when material undergoes a
phase transition.  An important consideration in practice is to assess
by how much the temperature can deviate from the correct solution if
the phase transition is not properly resolved. Our experiments are
therefore targeted at estimating how many mesh cells across a phase
transition are required to accurately model the temperature change.

\begin{figure}
  \centering
  \includegraphics[width=6cm]{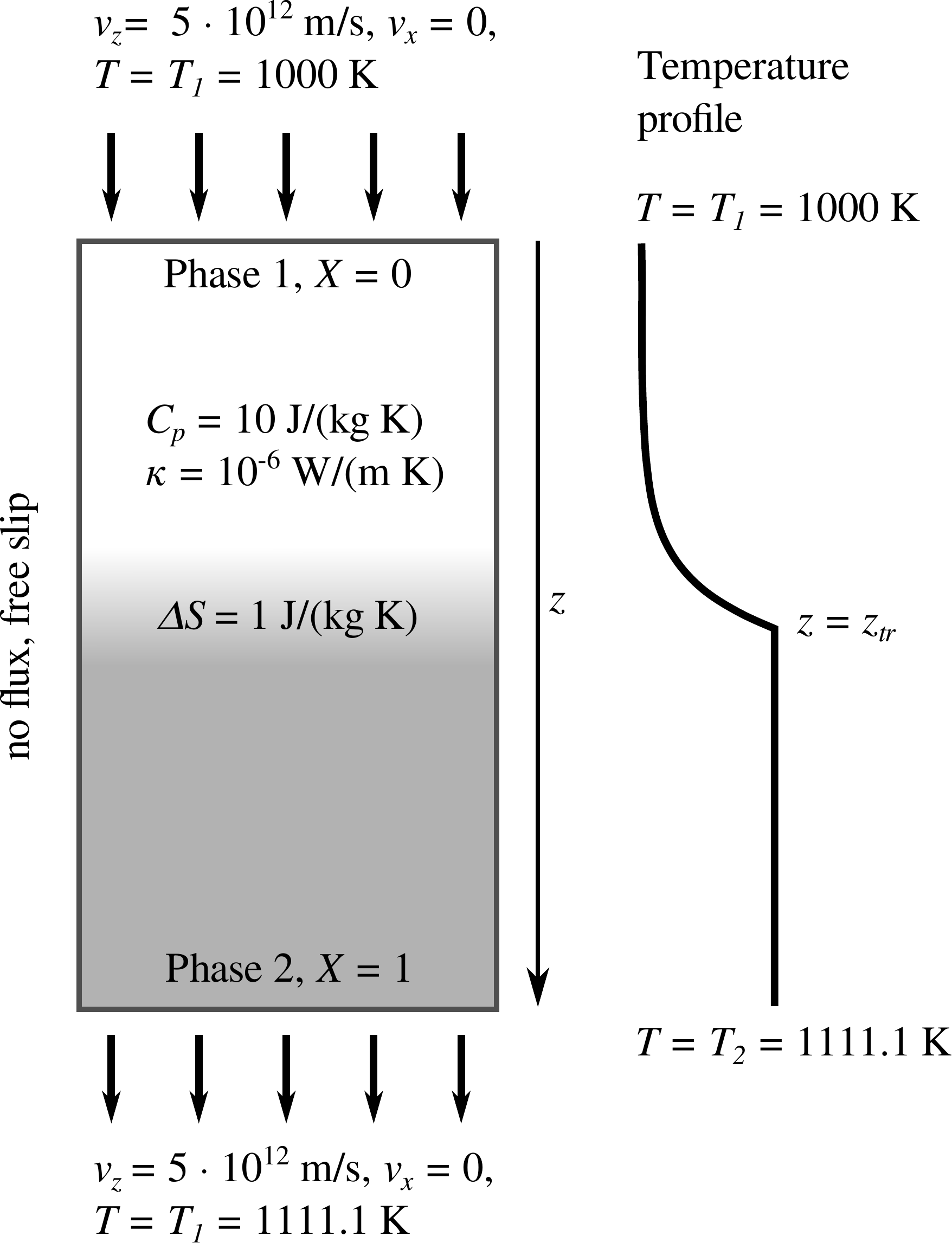}
    \caption{\it Setup of the latent heat benchmark together with the expected
      temperature profile across the phase transition. Material flows in with a prescribed
      temperature and velocity at the top, crosses the phase transition in the center and flows
      out at the bottom. }
  \label{fig:latent-heat-setup}
\end{figure}

The basic setup is a pipe with prescribed material inflow at constant velocity and temperature at the top, outflow at the bottom, and a univariant phase transition (occurring at a single value of temperature and pressure) approximately in the center of the domain (Fig.~\ref{fig:latent-heat-setup}). As initial condition, the model uses a uniform temperature field; however, when material crosses the phase transition, latent heat is released. 
In the steady state limit,
this leads to a temperature profile with a higher temperature in the
bottom half of the domain, which can be calculated by solving the
energy equation (equation~\eqref{eq:temp-latent-heat}, using approach (i)
above) for one-dimensional downward flow with (constant) vertical velocity $v_z$:
\begin{gather*}
\rho C_p
v_z
\frac{\partial T}{\partial z} = 
\rho T \Delta S v_z \frac{\partial X}{\partial z} 
+ \rho C_p \kappa
\frac{\partial^2 T}{\partial z^2}.
\end{gather*}
Here, $\rho C_p \kappa = k$ with $k$ the thermal conductivity and $\kappa$ the
thermal diffusivity. The latent heat generation is the product of the temperature $T$, the
entropy change $\Delta S$ across the phase transition divided by the specific
heat capacity and the derivative of the phase function $X$, which indicates the 
fraction of material transitioned from phase 1 to phase 2. If the velocity is smaller 
than a critical value (see also \cite{STO01} part 1, pp. 193--195), this latent heat term will be zero everywhere except for the one 
depth $z_\text{tr}$ where the phase transition occurs
discontinuously.

This means that there are two one-phase 
regions, one above $z_\text{tr}$ with only phase 1, and one below $z_\text{tr}$ with only 
phase 2, where the equation above (using the boundary conditions $T=T_1$ for $z \rightarrow -\infty $ 
and $T=T_2$ for $z \rightarrow \infty $) can be solved as
\begin{align*}
T(z) =\begin{cases}
T_1 + (T_2-T_1) e^\frac{v_z (z-z_\text{tr})}{\kappa}, & z<z_\text{tr},\\
T_2, & z>z_\text{tr}.
\end{cases}
\end{align*}
As we consider only the steady state, and the solution given  above tells us that for $z>z_\text{tr}$ 
(the region downward of the phase transition) the temperature is constant (see also the temperature profile 
in Fig.~\ref{fig:latent-heat-setup}), there is no net downward transport of heat from the phase 
change interface. In other words, the amount of heat generated at the phase transition is the same as 
the heat conducted upwards from the transition:
\begin{gather*}
\left.\rho v_z T \Delta S \right\rvert_{z=z_{\text{tr}^-}} 
= \frac{\kappa}{\rho C_p} \left.\frac{\partial T}{\partial z} \right\rvert_{z=z_{\text{tr}^-}} 
= \rho C_p v_z (T_2-T_1).
\end{gather*}
Rearranging this equation and using $T(z_\text{tr}) = T_2$ gives
\begin{gather*}
T_2 = \frac{T_1}{1 - \frac{\Delta S}{C_p}}.
\end{gather*}

In the numerical model, we can not exactly reproduce the behaviour of a Dirac delta
function as would result from taking the derivative $\frac{\partial X}{\partial z}$ of the
discontinuous phase function $X(z)$ that is considered in the benchmark. Rather, we use a hyperbolic 
tangent with a (small) finite width to model $X(z)$. This leads to a deviation of the numerical  
from the analytical solution that is dependent
on how well the mesh resolves the transition zone and how large one chooses the transition zone width to be. 
Both the mesh size and the width of the transition zone can be chosen
independently for numerical purposes.

\begin{figure}[ht]
  \includegraphics[width=0.48\textwidth]{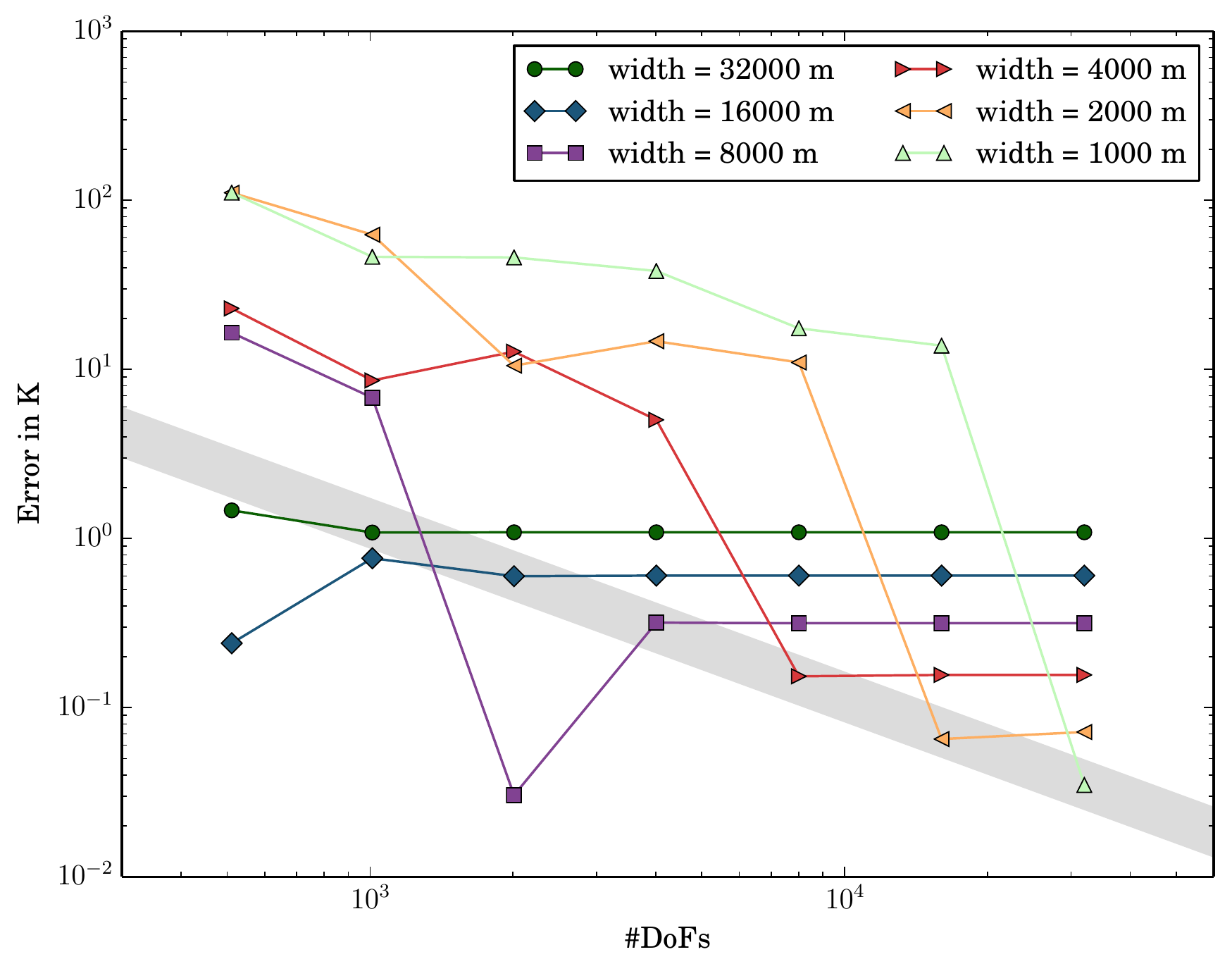}
  \caption{\it Results of the latent heat benchmark: Error of the modeled temperature $T_2$ at the 
               bottom of the model domain as a function of mesh resolution for different widths of 
               the phase transition. If the resolution is too low to resolve the phase transition, 
               errors are large ($>5$\,K) and do not vary in a systematic way, as grid points lie on random 
               points on the phase transition (or even exclude the phase transition).  
               If the phase transition is at least 4 cells wide (the gray area indicates models with exactly
               4 cells across the phase transition width), it is resolved properly and errors are 
               much smaller ($<5$\,K). In this case, the error mainly depends on the width of the phase transition 
               and converges for the width going to zero.\newline  
               The ``outlying'' blue and purple data points with unexpectedly small error result from models where the temperature 
               change across the phase transition was larger than the analytically predicted one (instead of 
               smaller, as for all the other models), and hence are by chance closer to the analytical solution.}
  \label{fig:latent-heat-results}
\end{figure}

Fig.~\ref{fig:latent-heat-results} shows numerical results that
demonstrate this interplay: If the resolution is high enough to
resolve the phase boundary (which requires approximately 4 mesh cells
across the phase transition, using bi-quadratic
finite elements, in our experiments), the error is small and is dominated by
the phase transition width -- the deviation of the approximate, smoothed model from the
exact one. On the other hand, while the mesh is too coarse to
resolve the transition zone, neither mesh refinement nor reducing
the size of the transition zone have a significant effect.

Hence, for modeling discontinuous phase transitions (or phase transitions 
that are too narrow to be resolved in the numerical model), 
to reach the highest accuracy the phase transition width
should be chosen as approximately four times of the smallest cell
size. This corresponds to the first data point after the kink of each
line in Fig.~\ref{fig:latent-heat-results}, i.e. the area highlighted in gray,
thus demonstrating predictable convergence.

\subsection{Mesh refinement}
\label{sec:mesh-refinement}

Many finite element
codes supporting adaptive mesh refinement use the ``Kelly''
refinement criterion \cite{GKZB83} to refine and coarsen the mesh in
response to the computed solution (for an overview of other error indicators 
used in computational geodynamics simulations, we refer to 
\cite{may2013overview,burstedde2013large,davies2011fluidity}). 
In the case of time-dependent
problems such as the one discussed here, one would perform this
adaptation every few time steps. The ``Kelly'' criterion computes a numerical
approximation to the second derivative of a finite element function
$v_h$, times a power of the mesh size, by evaluating for every cell $K$ the quantity
\begin{align*}
  \eta_K = 
  \left(
  h_K \int_{\partial K} |[\mathbf n \cdot \nabla v_h]|^2 \; dx
  \right)^{1/2},
\end{align*}
where $[\cdot]$ denotes the jump of the enclosed quantity across the
interface between cell $K$ and its neighbours, $\mathbf n$ is the
normal vector to the boundary of cell $K$, and $h_K$ is the
diameter of $K$.

This criterion was originally developed as an
error estimator for the Laplace equation \cite{KGZB83}, but
has been found widely useful in adaptive meshing because it also estimates the
polynomial interpolation error on every cell. It has thus been used
for many different equations to generate good meshes, even if no
provably accurate error estimators are available for these equations. 

In the context of mantle convection, it therefore seems appropriate to
drive mesh refinement by applying this criterion to either the
temperature or velocity field. Indeed, we advocated for this approach in
\cite{KHB12} based on the observation that this should help reduce the
error in the natural energy norms for these two solution variables.

On the other hand, in actual applications, one is often interested in
a variety of quantities that 
are, at best, tangentially related to the energy norm error and whose
approximation is not always improved by choosing a mesh based on an
energy error indicator. A typical example would be simulations that
investigate the importance of phase changes on the dynamics
of convection: While the coefficients in the equations (e.g., density, viscosity) and possibly other derived
quantities such as seismic velocities are discontinuous at these interfaces, the solution fields
(e.g., temperature and
velocity) may vary in ways that do not make such interfaces
obvious. Consequently, only refining based on velocity and temperature
may not yield meshes that reveal these phase boundaries in sufficient
detail to really capture their small-scale influences. Furthermore,
the meshes so generated would not allow to extract interfaces with
sufficient resolution to account for the dynamic effects of phase changes, 
latent heat transfer as
discussed in Section~\ref{sec:latentheat}, or for comparison 
against observations like seismic tomographic models.

\subsubsection{A practical approach}

Despite the fact that we have well over a decade of experience with mesh adaptation algorithms,
it is not clear to us how one can devise
methods that \textit{automatically} take into account what one may be
interested in. Dual Weighted Residual methods such as those discussed
in \cite{BR03} may be appropriate but are unwieldy to implement for
time-dependent problems. Instead, the best solution we can come up with
is a complex but flexible, two-tiered
system for adaptive mesh refinement that is primarily driven by
letting users choose what information they think is most important for
their purposes. In a first step, we compute refinement indicators
$\eta_K^{(1)},\ldots,\eta_K^{(L)}$ by choosing among a
list of indicators that include the following:
\begin{itemize}
\item The ``Kelly'' indicator applied to the velocity or temperature.
\item A weighted discrete approximation of the gradient,
  \begin{align*}
    \eta_K = 
    h_K^{1+d/2} |\nabla_h v_h(\mathbf x_K)|.
  \end{align*}
  Here, $\mathbf x_K$ is the center of $K$, $d$ the space dimension,
  and the factor $h_K^{1+d/2}$ is chosen so that indicators converge
  to zero as the mesh size $h\rightarrow 0$ even for discontinuous
  discretizations $v_h$ of otherwise continuous exact solutions $v$. This criterion is then
  applied to derived quantities $v_h$ such as the density, the viscosity, or
  the thermal energy density $\rho C_p T$.
\end{itemize}
The criteria $\eta_K^{(\ell)}$ are then scaled or normalized to
yield $\tilde\eta_K^{(\ell)}$,
and the final refinement indicators are obtained by either computing
the maximum of (scaled or normalized) error indicators, $\eta_K=\max_{1\le\ell\le L}
\tilde\eta_K^{(\ell)}$, or the sum of these indicators,
$\eta_K=\sum_{1\le\ell\le L} \tilde\eta_K^{(\ell)}$. Cells are then marked
for coarsening or refinement based on $\eta_K$.

There are also cases where refinement needs to be driven
algorithmically, rather than based on criteria derived from solution
or derived values. For example, we have found that it is often useful
to only refine in a region of particular interest, even though the
model is larger; in these cases, one can think of the larger model
(with a relatively coarse mesh) as providing self-consistent boundary
values for the smaller region of interest (with a finer mesh). Another
example is to ensure a minimal refinement level for all cells at the
surface, or at a particular depth.

This approach provides great flexibility in defining how and
where the mesh is refined, as necessary, and thereby provide high
accuracy where it is important for the particular question one wants to
investigate in a simulation. At the same time, there is little
theoretical underpinning that this approach is ``optimal'' (however
one may want to define this).

\subsubsection{Mesh refinement in 2-D spherical convection}
\label{sec:results-mesh-refinement}

We demonstrate the flexibility provided by the mesh refinement
procedure using an
example of 2D mantle convection that includes phase transitions and
the associated discontinuities of density and viscosity. The geometry
is a spherical shell, and the mantle is heated from the bottom, where
the temperature is fixed to 2600\,K, and cooled from the top, where
the temperature is 273\,K. No additional heating processes (such as
shear heating, adiabatic heating, or latent heat) are included, and
the initial temperature is constant at 1600\,K except for the two
thermal boundary layers.  

We model two phase transitions at depths of 410\,km and
660\,km (reflecting the olivine-spinel and
spinel-perovskite transformations), where
both viscosity and density change discontinuously. Specifically,
we use a viscosity
\begin{align}
  \eta = \eta_0 e^{-E \frac{T - T_\text{ref}}{T_\text{ref}}}.
\end{align}
where $\eta_0 = 10^{21}$\,Pa\,s in the upper mantle,
$\eta_0=10^{22}$\,Pa\,s between 410\,km and 660\,km depth, and
$\eta_0=10^{23}$\,Pa\,s in the lower mantle; we choose
the dimensionless activation energy $E = 15$, and the reference 
temperature $T_{\text{ref}} = 1600$\,K.

Our density model satisfies
\begin{align}
  \rho = & \rho_0 (1 + \kappa p) (1 - \alpha (T - T_\text{ref}))\\
  \notag
  &+ \begin{cases} 0, & \text{depth} < \text{410\,km}\\
                   \Delta\rho_{410}, & \text{410\,km} < \text{depth} < \text{660\,km}\\
                   \Delta\rho_{410} + \Delta\rho_{660}, & \text{depth} > \text{660\,km} \end{cases}.
\end{align}
with $\rho_0 = 3300$\,kg/m\textsuperscript{3}, 
$\kappa = 5.124 \cdot 10^{12}$\,Pa\textsuperscript{-1}, 
$\alpha = 4 \cdot 10^{-5}$\,K\textsuperscript{-1},
and density increases
of $\Delta\rho_{410} = 100$\,kg/m\textsuperscript{3} and 
$\Delta\rho_{660} = 200$\,kg/m\textsuperscript{3} at
the 410\,km and 660\,km phase transitions.  Velocities at the outer
boundary are prescribed, using the present-day plate velocities along
the equator projected onto the two-dimensional slice used in our model
\cite{gurnis2012plate}.

\begin{figure}
 \centering
 \vspace{-1em}
 \includegraphics[width=0.31\textwidth]{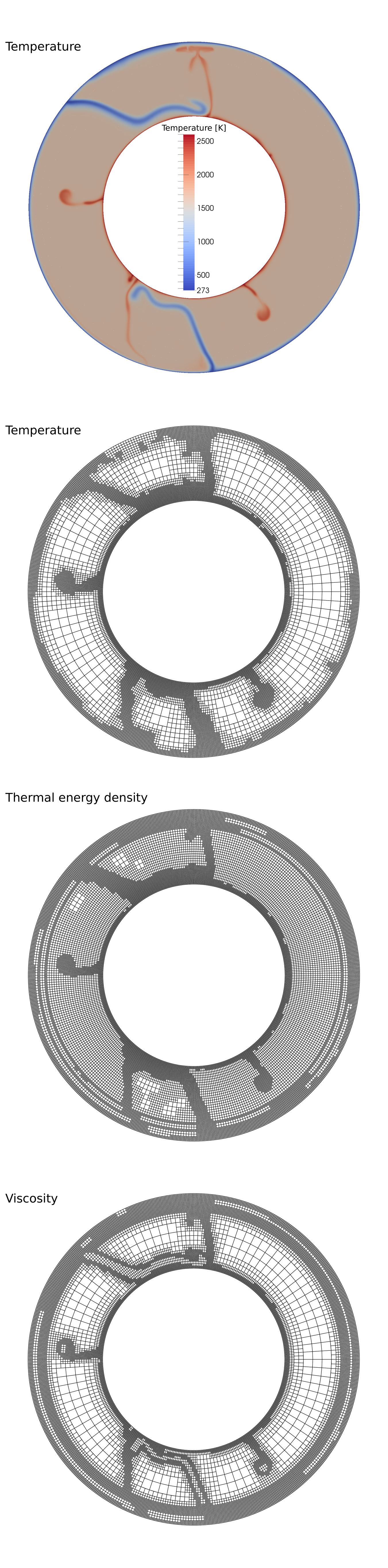}
 \vspace{-2em}
 \caption{Temperature distribution and mesh in a 2D mantle convection model, using different refinement criteria:
          The Kelly error estimator for the temperature field,
          an error indicator based on the magnitude of the approximate gradient of the thermal energy density $\rho C_p T$,
          and the approximate gradient of the viscosity.}
 \label{fig:mesh_refinement}
\end{figure}

Fig.~\ref{fig:mesh_refinement} shows the temperature distribution in
this model after 260 million years, together with the
corresponding meshes generated using different criteria for the
adaptive refinement. The figure illustrates how a refinement criterion
based solely on the temperature almost entirely misses the phase
transitions in favour of resolving only the boundary layers and plumes. It would
therefore not yield sufficiently resolved fields 
for comparisons with tomographic models of Earth. On the other hand,
refining based on weighted approximate gradients of either the
thermal energy density $\rho C_p T$ or the viscosity $\eta$ allows the
resolution of phase boundaries.

Which of these meshes yields the ``best'' solution cannot be
quantified without specifying what the ``goal'' of the simulation
is. It is possible that the meshes refined based on the thermal energy
density or the density have a larger energy norm error in the velocity
and/or temperature. On the other hand, their accuracy in predicting
tomographically visible interfaces is certainly much higher.

\subsection{Tracking chemical compositions and other quantities}
\label{sec:composition}

In many complex simulations of mantle convection, it is necessary to
track not only the flow of thermal energy (described by
equation~\eqref{eq:boussinesq-3}), but also how the chemical
composition, trace or radiogenic elements, isotope ratios, water content
-- or other quantities such as grain sizes -- are transported along
with the velocity.

In mantle convection codes, this has traditionally (and successfully)
be done using tracer particles \cite{poliakov1992,gerya2003,McNamara2004,Popov2008,TMK14}. However, it is not trivial to
implement tracers efficiently and scalably in the context of
large-scale parallel codes with dynamically changing, adaptively
refined meshes, as opposed to globally refined, statically
partitioned, fixed meshes. A number of these
challenges -- and possible solutions -- are discussed in more
details in \cite{Gassmoeller2016}.

On the other hand, many of the \textit{applications}
that have traditionally motivated the use of particles can equally
well be done by using a field-based description of the quantities one
wants to advect along. The advantage in using this approach is the
well developed numerical infrastructure for solving advection
equations, and the ease
with which these can then be evaluated at 
quadrature points when computing material properties; highly
efficient tools are also available in many of the available finite element libraries
to facilitate data movement upon mesh refinement and
repartitioning (see, for example, \cite{BBHK10}).

Using field-based approaches then requires advecting any number of
``compositional fields'' $C_i$ along with the velocity field, by solving
the advection equations
\begin{align}
  \label{eq:composition}
  \frac{\partial C_i}{\partial t} + \ve u \cdot \nabla C_i & = Q_i \qquad \text{for $i=1...n$},
\end{align}
where $Q_i$ are source terms that may depend on velocity, pressure,
temperature, and the compositional fields $C_i$ themselves. Through
appropriate choices of these source terms, one can also
model reactions among fields, for example to describe compositional
changes upon partial melting or freezing of material. On the other hand, entirely different
reactions can equally easily be modelled, and we will outline one
example in Section~\ref{sec:results-composition} below.

In practice, the compositional fields are easily evaluated at
quadrature points, and can therefore be used to affect the description
of material parameters such as the density and viscosity.

\subsubsection{Implementation}

As stated, equation~\eqref{eq:composition} does not contain any
diffusion, in line with the fact that chemical species do not diffuse
at appreciable rates on length scales of the Earth mantle. Consequently, the
numerical solution of \eqref{eq:composition} presents challenges when modeling sharp gradients
-- for example, when tracking chemical heterogeneities, or when using
the fields $C_i$ to track where material that originates from one
particular area is transported over time. To stabilize the numerical
solution, one typically employs one of many artificial viscosity
schemes, such as the SUPG formulation \cite{brooks1981petrov, brooks1982streamline}
or schemes based on the
residual of an entropy equation \cite{GPP11,KHB12}. This is of course
also necessary for the temperature equation~\eqref{eq:boussinesq-3}.

In practical applications, it has proven useful to allow descriptions
of the source terms $Q_i$ that may consist both of finite but time
dependent components, and of impulse functions in time. An example for
the use of impulse functions is where the $C_i$ describe the chemical
composition of rocks; if these compositions change due to partial melting 
and melt extraction that happen instantaneously (compared to the size of a time
step) as a rock moves through the $p-T$ phase diagram, the
compositions $C_i$ also need to change instantaneously, rather than
continuously. Allowing both continuous and impulse components can be
achieved by providing the functions that compute $Q_i$ with current
values of strain rate, temperature, pressure, compositions, and
spatial location, along with a time increment $\Delta t$, and require
them to
return $\int_t^{t+\Delta t} Q_i(\varepsilon(\mathbf u),p,T,C,\mathbf
x,\tau)\; d\tau$. If $Q_i$ contains impulse components, then the
functions' return value will
simply have a contribution that is not proportional to $\Delta t$.

\subsubsection{Tracking finite strain}
\label{sec:results-composition}

We demonstrate the flexibility of using compositional fields using the
example of a Cartesian convection model that tracks the accumulated
finite strain at every location of the domain. 
For this purpose, we define $C_i$ as the components of the deformation gradient (or deformation) tensor $\boldsymbol{F}$, 
which represents the deformation accumulated over time by idealized little grains of finite size.
This is done in such a way that (in 2D) $C_1 = \boldsymbol{F_{xx}}$, $C_2 = \boldsymbol{F_{xy}}$, etc.
The time derivative of $\boldsymbol{F}$ can be computed as
 \begin{align}
  \label{eq:finite_strain}
\frac{\partial \boldsymbol{F}}{\partial t} = \boldsymbol{G} \boldsymbol{F},
\end{align} 
where $\boldsymbol{G} = \nabla \mathbf u^T$ is the velocity gradient tensor
\cite{McKenzie1983, dahlen1998theoretical, Becker2003}.
The initial deformation is $\boldsymbol{F_0} = \boldsymbol{I}$, with $\boldsymbol{I}$ being the identity tensor. 

This means that the $Q_i$ on the right-hand side of Equation \eqref{eq:composition} can be computed as the
product of the current velocity gradient $\boldsymbol{G}$ and the 
accumulated deformation $\boldsymbol{F}$ at the previous time step.

A direct visualization of $\boldsymbol{F}$ is not intuitive, because it contains rotational components that represent a rigid body rotation without deformation. Following \cite{Becker2003} we can polar-decompose the tensor into a positive-definite and symmetric tensor $\boldsymbol{L}$, and an orthogonal rotation tensor $\boldsymbol{R}$, as $\boldsymbol{F} = \boldsymbol{L} \boldsymbol{R}$, therefore $\boldsymbol{L}^2 = \boldsymbol{L}\boldsymbol{L}^T = \boldsymbol{F} \boldsymbol{F}^T$. The left stretching tensor $\boldsymbol{L}$ then describes the deformation we are interested in, and its eigenvalues $\lambda_i$ and eigenvectors $\boldsymbol{e}_i$ describe the length and orientation of the half-axes of the finite strain ellipsoid. Moreover, we will represent the amount of relative stretching at every point by the ratio $\ln(\lambda_1/\lambda_2)$, called the \textit{natural strain} \cite{Ribe1992}.

\begin{figure}
 \centering
 \includegraphics[width=0.48\textwidth]{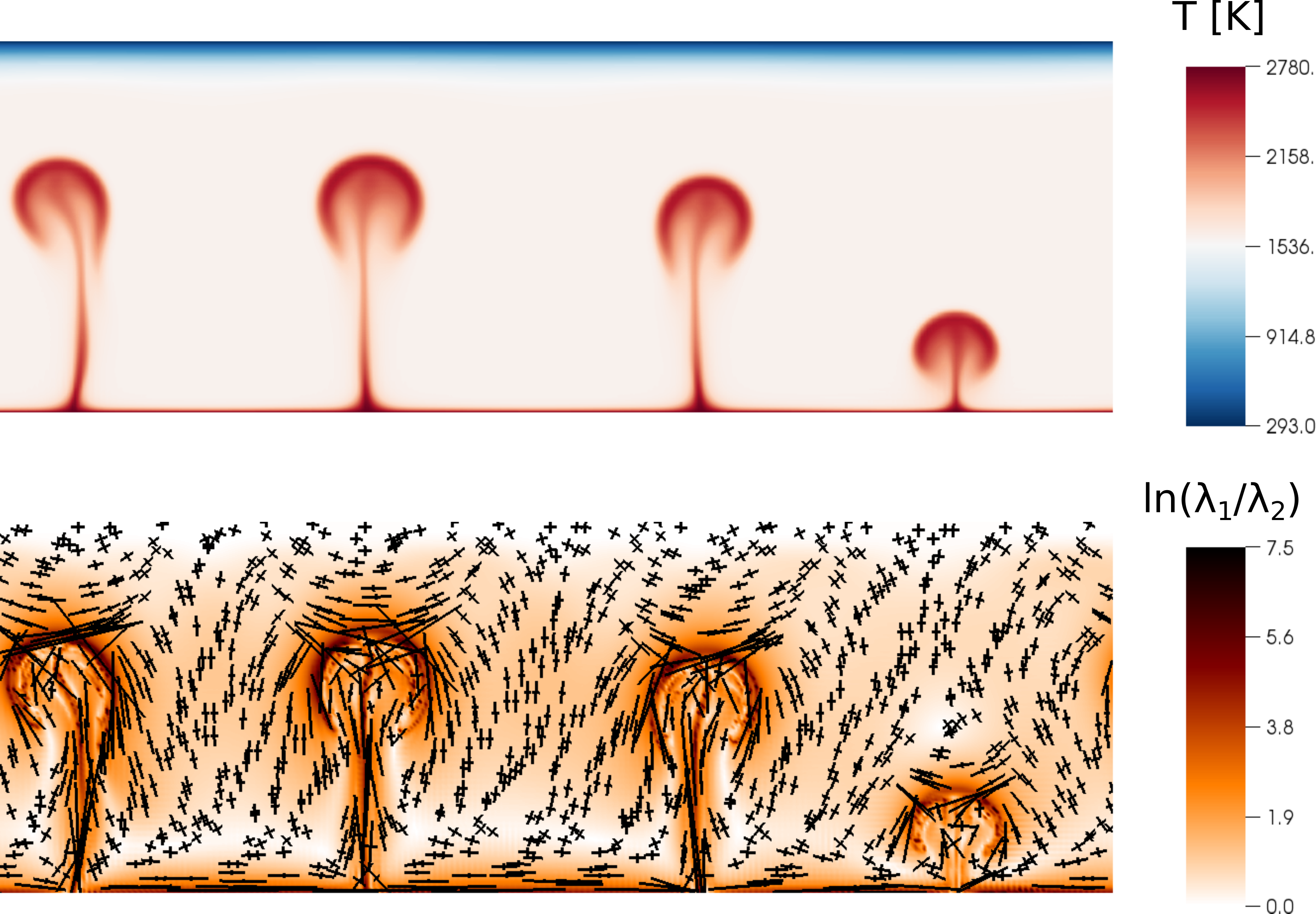}
 \caption{Temperature (top) and accumulated natural strain (bottom) in a 2D Cartesian convection model at a time of 67.6~Ma. 
          Black crosses represent the scaled eigenvectors of the stretching tensor $\boldsymbol{L}$, 
          showing the direction of stretching and compression the material has experienced.}
 \label{fig:finite_strain}
\end{figure}

The model we present here as an example for tracking of finite strain features a 
box with an aspect ratio of three and dimensions of 2900 $\times$ 8700\,km. The mantle is cooled from the top (where the 
temperature is 293\,K) and heated from the bottom (where the temperature is 2780\,K) with no additional heat sources in 
the form of internal heating or latent heat. 
The density is modelled as 
\begin{align}
  \rho = \rho_0 \left(1 - \alpha (T - T_\text{ref}) \right), 
\end{align}
with $\rho_0 = 3400$~kg~m\textsuperscript{-3}, $\alpha = 2 \times 10^{-5}$~K\textsuperscript{-1} and the reference temperature $T_\text{ref} = 1600$~K. 
Thermal conductivity and gravity are set to $k=4.7$~W~m\textsuperscript{-1}~K\textsuperscript{-1} and $g=9.81$~m~s\textsuperscript{-2}.
We choose the temperature-dependent viscosity as
\begin{align}
  \eta = \eta_0 e^{-E \frac{T - T_\text{ref}}{T_\text{ref}}}, 
\end{align}
with $\eta_0 = 5 \times 10^{21}$~Pa~s and $E=7$.
Hence, the bottom thermal boundary layer, where viscosities are lower, becomes unstable first, and plumes start to rise 
towards the surface, see Fig.~\ref{fig:finite_strain} (top).
Material moves to the sides at the top of the plume head, so that it is shortened in vertical direction (short black vertical lines in Fig.~\ref{fig:finite_strain}, bottom) and stretched in horizontal direction (long horizontal lines). The sides of the plume head 
show the opposite effect. Shear occurs mostly at the edges of the plume head, in the plume tail, and in the bottom boundary layer 
(black areas in the natural strain distribution).

\section{Application to a complex problem}
\label{sec:application}

In order to present the methods discussed in the previous sections in
practice, we here show results of a global mantle convection model
that combines a compressible formulation 
with earth-like material properties, a strongly temperature dependent viscosity, 
chemical heterogeneities tracked by compositional fields,
and prescribed surface velocities.

In particular, the model geometry resembles Earth's mantle, and starts
from an undisturbed, motionless state. A layer of dense basaltic
material with initially uniform thickness of $150$~km covers the core-mantle boundary, and
the initial temperature profile follows an adiabat of $1613$~K
computed with the material properties that are provided by the Perple\_X 
software \cite{Connolly2005} based on a database of mineral properties 
\cite{Stixrude2011}, overall a method similar to \cite{NTDC09}. 
This approach yields realistic, earth-like material properties, but
also entails several challenges, such as discrete sampling in
pressure--temperature space, and quasi-discontinuous jumps due to phase
transitions. The viscosity is based on a published viscosity model
incorporating constraints from mineral physics, geoid deformation and
seismic tomography \cite{ste06a}. It is depth- and temperature
dependent with a depth-dependent activation enthalpy of $200$--$500$~kJ/mol
and would lead to viscosity variations of at least eight orders of
magnitude over the model temperature range. 
In order to limit the maximal velocity, and thus the number of timesteps
and computational cost, we artificially restrict
the viscosity to the range $5 \times 10^{19}$--$1.5 \times 10^{23}$~Pa~s 
by cutting off values outside of this range. 
As has been shown elsewhere, see \cite{dannberg2016compressible,tosi2015community},
our Stokes solver is capable of solving larger viscosity 
contrasts up to at least seven orders of magnitude.
Surface velocities in the model are prescribed using published plate
reconstructions \cite{Seton2012}, and are prepared
by the GPlates software \cite{Boyden2011} at discrete positions, and
interpolated to the adaptively refined mesh within \aspect{}. Boundary
temperatures are prescribed to $273$~K (at the surface) and $3700$~K
(at the core-mantle boundary). The resolution of the finest mesh cells
is 23 km (large portions of the model are adaptively coarsened), and
the overall computation has about 100 million degrees of freedom in
each time step. The model requires 3,700
time steps, within which we iterate out the nonlinearity with on
average about 2 sub-iterations (for a total of 6,000 
nonlinear iterations). The model required a computing time of
24.5 hours on 1536 processes, i.e., 37,600 CPU hours.

\begin{figure}
 \centering
 \includegraphics[width=0.48\textwidth]{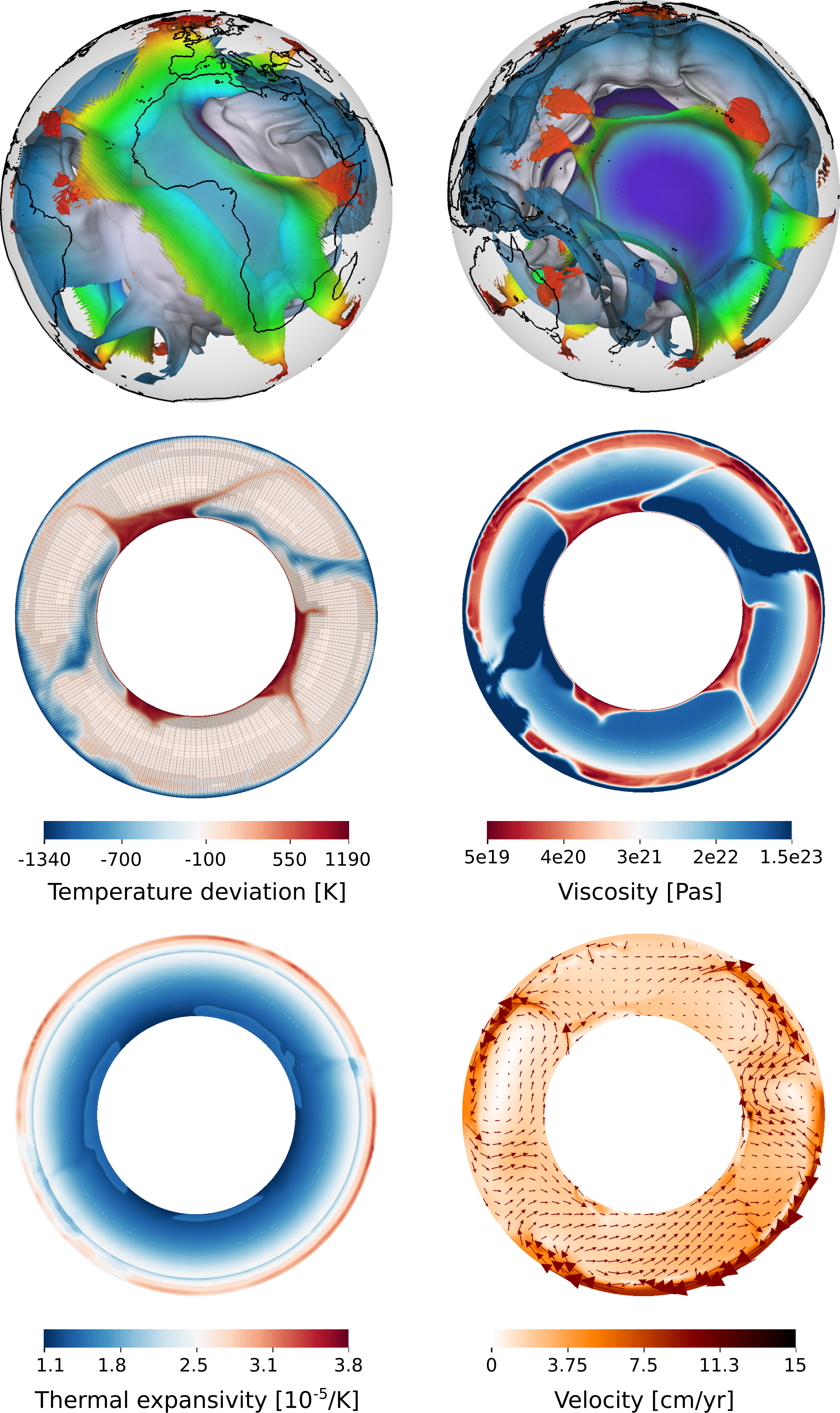}
 \caption{Final state of a global mantle convection simulation after
   250 Ma of model time. Top panels: Isosurfaces of -150~K (white to
   blue) and +300~K (rainbow colored) temperature deviation from an
   adiabatic temperature profile for the African hemisphere (left) and
   the Pacific hemisphere (right). Colors visualize height above the
   core-mantle boundary, and coastlines are shown in black
   outlines. Center and bottom panels: Equatorial slices through the model showing temperature deviation and finite-element mesh (center left), viscosity (center right), thermal expansivity (bottom left), and velocity (bottom right). In all slices the Greenwich meridian is ``up'' and the view is directed from the North pole to the South pole.}
 \label{fig:application}
\end{figure}

The model results presented in Fig.~\ref{fig:application} demonstrate the
complexities that arise in realistic mantle convection models as discussed in this paper. The
strongly temperature dependent viscosity leads to narrow
upwelling plumes (with diameters around 100~km) that rise from the edges of the dense basal piles and
reach the surface close to observed hotspot locations on Earth as
observed in many other studies \cite{steinberger2012geodynamic,davies2012reconciling,bower2013lower,hassan2015provenance}.
Due to its low viscosity the plume material moves with velocities larger than $10$~cm/yr in the upper mantle thus limiting the timestep length of the model.
The mineral physics based material properties contain sharp gradients
in density, thermal expansivity, and specific heat capacity; these are
particularly prominent in the mantle transition zone and at the
Bridgmanite-Postperovskite transition close to the core-mantle
boundary in the lower left panel of Fig.~\ref{fig:application}.

The more complex setup of our model compared to earlier studies --
including compressibility, highly temperature-dependent viscosity, and
more complex material parameters -- does not change the basic results
of the computation. However, the setup focuses the plumes into
narrower structures, and the higher accuracy possible with our methods
allows additional use cases for the model results:
Velocities, temperature and compositions can be used as constraints
for regional high-resolution models investigating particular processes
such as the interaction between rising mantle plumes and mid-ocean
ridges \cite{gassmoller2016major}, or the generation and ascent of
chemically zoned plumes that are thought to be responsible for the
generation of zoned hotspot tracks \cite{weis2011role}. Consistent
temperature and pressure profiles of compressible models also allow for a more straightforward comparison between geodynamic and seismic models, for example by converting the geodynamic model results to a synthetic tomography model\cite{ritsema2007tomographic}, or by using the created seismic velocity field to forward-model seismic wave propagation \cite{nissen2014}.

\section{Conclusions}
\label{sec:conclusions}

Mantle convection codes have provided a great deal of insight into the
dynamics of the mantles of Earth and other rocky planets. Yet, to
deepen their veracity requires both increasing the complexity of the
models they solve (e.g., in dealing with highly variable coefficients
and latent heat), as well as the scale at which they can discretize
these models for a computational solution (e.g., in devising
adaptively refined meshes). Both of these challenges require
going beyond the ways in which most codes have so far operated.

In this contribution, we have summarized some of the lessons we have
learned over the past years in solving complex mantle convection
problems using state-of-the-art computational methods. Specifically,
we have discussed effective ways for dealing with time stepping,
compressibility, discontinuous coefficients, latent heat, adaptively
refining finite element meshes, and advecting additional
quantities. None of these techniques by themselves are sufficient to
deal with the most complex models we have encountered, but jointly,
and in concert with the methods previously discussed in \cite{KHB12},
they help solve some of the most complex mantle convection models we
know of on large-scale compute clusters. We believe that they will
also be useful in using even more complicated models -- for example
with material models that utilize grain size evolution, track finite
strain, consider diffusion and dislocation creep, plasticity effects,
and other inputs -- to accurately predict outputs that can be compared
to available data via seismic imaging, surface heat fluxes, plate
velocities, and other measurements.

\section{Acknowledgements}

The authors would like to thank Scott King and Eh Tan for their help 
with reproducing the benchmark of \cite{KLVLZTTK10}. Cedric Thieulot provided
the initial motivation to consider averaging of material parameters in Section~\ref{sec:averaging}.

All authors were partially supported by the Computational
Infrastructure for Geodynamics initiative (CIG), through the National
Science Foundation under Awards No.~EAR-0949446 and EAR-1550901,
administered by The University of
California-Davis. TH was partially supported by the National Science Foundation grant
DMS-1522191. JD, RG, and WB were partially supported by the National Science
Foundation under award OCI-1148116 as part of the Software Infrastructure for
Sustained Innovation (SI2) program.

The compute time for the computations shown in Section~\ref{sec:application} was provided by the North-German Supercomputing Alliance (HLRN) as part of the project bbk00003 ``Numerical Geodynamics: Plume-Plate interaction in 3D mantle flow -- Revealing the role of internal plume dynamics on global hot spot volcanism''.

Clemson University is acknowledged for generous allotment of compute
time on the Palmetto cluster.

The authors greatly appreciate all of these sources of support.

\bibliographystyle{gji}
\bibliography{paper}

\appendix
\newpage
\section{King results}

Given how widely used the benchmark defined in \cite{KLVLZTTK10} is,
Tables~\ref{fig:king_full_results_ala} and
\ref{fig:king_full_results_tala} provide a full account of our results
for this benchmark using the strategy to solve compressible equations
discussed in Section~\ref{sec:compressibility}. In particular, the
tables show convergence as the mesh size goes to zero, and
extrapolated values that can be compared against the values that were
reported in \cite{KLVLZTTK10}.

\begin{table*}
\centering
\begin{tabular}{llllllll}
  Di & Ra & 1/h & Nu & Vrms & $\left<T\right>$ & $\phi$ & W \\ \hline
  0.25 & $10^4$ & 16 & 4.53819 & 40.02007 & 0.51514 & 0.85213 & 0.85157 \\
  0.25 & $10^4$ & 32 & 4.45192 & 39.96357 & 0.51496 & 0.84984 & 0.84928 \\
  0.25 & $10^4$ & 64 & 4.42482 & 39.95753 & 0.51494 & 0.84960 & 0.84903 \\
  0.25 & $10^4$ & 128 & 4.41735 & 39.95684 & 0.51494 & 0.84957 & 0.84901 \\
  \rowcolor{Gray} \rule{0pt}{2.2ex}
  &  & extrapolated & 4.41450 & 39.95676 & 0.51494 & 0.84957 & 0.84900 \\
  0.25 & $10^4$ & King UM & 4.406 & 39.952 & 0.515 & 0.847 & 0.849 \\
  0.25 & $10^4$ & King VT & 4.4144 & 40.0951 & 0.5146 & 0.849 & 0.849 \\
  0.25 & $10^4$ & King CU & 4.41 & 40 & 0.5148 & 0.8494 & 0.8501 \\ 
  \hline
  0.5 & $10^4$ & 16 & 3.91228 & 35.98789 & 0.52271 & 1.38719 & 1.38541 \\
  0.5 & $10^4$ & 32 & 3.84891 & 35.94470 & 0.52245 & 1.38402 & 1.38225 \\
  0.5 & $10^4$ & 64 & 3.82932 & 35.93997 & 0.52241 & 1.38368 & 1.38190 \\
  0.5 & $10^4$ & 128 & 3.82399 & 35.93943 & 0.52241 & 1.38364 & 1.38187 \\
  \rowcolor{Gray} \rule{0pt}{2.2ex}
   &  & extrapolated & 3.82200 & 35.93936 & 0.52241 & 1.38363 & 1.38186 \\
  0.5 & $10^4$ & King UM & 3.812 & 35.936 & 0.522 & 1.381 & 1.381 \\
  0.5 & $10^4$ & King VT & 3.8218 & 36.0425 & 0.5214 & 1.3812 & 1.3812 \\
  0.5 & $10^4$ & King CU & 3.82 & 35.9 & 0.5217 & 1.3818 & 1.383 \\
  \hline
  1 & $10^4$ & 16 & 2.47804 & 24.69538 & 0.51160 & 1.34460 & 1.35568 \\
  1 & $10^4$ & 32 & 2.45507 & 24.68259 & 0.51145 & 1.34286 & 1.35415 \\
  1 & $10^4$ & 64 & 2.44835 & 24.68113 & 0.51143 & 1.34270 & 1.35399 \\
  1 & $10^4$ & 128 & 2.44659 & 24.68096 & 0.51142 & 1.34268 & 1.35398 \\
  \rowcolor{Gray} \rule{0pt}{2.2ex}
   &  & extrapolated & 2.44596 & 24.68094 & 0.51142 & 1.34268 & 1.35397 \\
  1 & $10^4$ & King UM & 2.438 & 24.663 & 0.512 & 1.343 & 1.349 \\
  1 & $10^4$ & King VT & 2.4716 & 25.0157 & 0.51 & 1.3622 & 1.3621 \\
  1 & $10^4$ & King CU & 2.47 & 24.9 & 0.5103 & 1.3627 & 1.3638 \\
  \hline
  0.25 & $10^5$ & 16 & 9.83522 & 179.93650 & 0.53284 & 2.09314 & 2.09246 \\
  0.25 & $10^5$ & 32 & 9.53887 & 178.40376 & 0.53247 & 2.05964 & 2.05880 \\
  0.25 & $10^5$ & 64 & 9.33472 & 178.11210 & 0.53220 & 2.05331 & 2.05248 \\
  0.25 & $10^5$ & 128 & 9.26701 & 178.07926 & 0.53216 & 2.05260 & 2.05177 \\
  \rowcolor{Gray} \rule{0pt}{2.2ex}
   &  & extrapolated & 9.23341 & 178.07510 & 0.53216 & 2.05251 & 2.05168 \\
  0.25 & $10^5$ & King UM & 9.196 & 178.229 & 0.532 & 2.041 & 2.051 \\
  0.25 & $10^5$ & King VT & 9.2428 & 179.7523 & 0.5318 & 2.0518 & 2.0519 \\
  0.25 & $10^5$ & King CU & 9.21 & 178.2 & 0.5319 & 2.0503 & 2.054 \\
  \hline
  0.5 & $10^5$ & 16 & 8.02846 & 156.42656 & 0.54891 & 3.28922 & 3.28780 \\
  0.5 & $10^5$ & 32 & 7.77804 & 155.33598 & 0.54847 & 3.24554 & 3.24386 \\
  0.5 & $10^5$ & 64 & 7.63386 & 155.14464 & 0.54809 & 3.23782 & 3.23615 \\
  0.5 & $10^5$ & 128 & 7.58838 & 155.12248 & 0.54805 & 3.23693 & 3.23526 \\
  \rowcolor{Gray} \rule{0pt}{2.2ex}
   &  & extrapolated & 7.56741 & 155.11957 & 0.54804 & 3.23682 & 3.23514 \\
  0.5 & $10^5$ & King UM & 7.532 & 155.304 & 0.548 & 3.221 & 3.233 \\
  0.5 & $10^5$ & King VT & 7.5719 & 156.5589 & 0.5472 & 3.2344 & 3.2346 \\
  0.5 & $10^5$ & King CU & 7.55 & 155.1 & 0.5472 & 3.233 & 3.2392 \\
  \hline
  1 & $10^5$ & 16 & 4.01908 & 84.62206 & 0.53004 & 2.77354 & 2.78862 \\
  1 & $10^5$ & 32 & 3.91951 & 84.38966 & 0.52998 & 2.75378 & 2.77104 \\
  1 & $10^5$ & 64 & 3.88354 & 84.37059 & 0.52983 & 2.75208 & 2.76937 \\
  1 & $10^5$ & 128 & 3.87364 & 84.36817 & 0.52981 & 2.75189 & 2.76918 \\
  \rowcolor{Gray} \rule{0pt}{2.2ex}
   &  & extrapolated & 3.86988 & 84.36782 & 0.52980 & 2.75187 & 2.76916 \\
  1 & $10^5$ & King UM & 3.857 & 84.587 & 0.53 & 2.742 & 2.765 \\
  1 & $10^5$ & King VT & 3.878 & 85.5803 & 0.5294 & 2.761 & 2.7614 \\
  1 & $10^5$ & King CU & 3.88 & 84.6 & 0.5294 & 2.7652 & 2.7742 \\
  \hline
  \end{tabular}

  \caption{Compressible results using the ALA formulation of convection
    corresponding to the benchmark defined in \cite{KLVLZTTK10}
    (see Section~\ref{sec:compressible-2d-cartesian}). The ASPECT results were obtained by running the benchmark on increasingly finer meshes, and extrapolating from the 1/128 mesh using Richardson extrapolation. Acronyms for the different codes are as in Table~\ref{table:king_results_short}.}
 \label{fig:king_full_results_ala}
\end{table*}

\begin{table*}
\centering
\begin{tabular}{llllllll}
  Di & Ra & 1/h & Nu & Vrms & $\left<T\right>$ & $\phi$ & W \\ \hline
  0.25 & $10^4$ & 16 & 4.54966 & 40.11121 & 0.51292 & 0.85535 & 0.85306 \\
  0.25 & $10^4$ & 32 & 4.46241 & 40.05425 & 0.51276 & 0.85304 & 0.85075 \\
  0.25 & $10^4$ & 64 & 4.43490 & 40.04816 & 0.51274 & 0.85279 & 0.85050 \\
  0.25 & $10^4$ & 128 & 4.42730 & 40.04747 & 0.51273 & 0.85277 & 0.85047 \\
  \rowcolor{Gray} \rule{0pt}{2.2ex}
   &  & extrapolated & 4.42440 & 40.04738 & 0.51273 & 0.85276 & 0.85047 \\
  0.25 & $10^4$ & King UM & 4.416 & 40.043 & 0.513 & 0.85 & 0.85 \\
  0.25 & $10^4$ & King VT & 4.43 & 40.2 & 0.5127 & 0.8535 & 0.851 \\
  0.25 & $10^4$ & King CU & 4.42 & 40.1 & 0.5129 & 0.8539 & 0.8521 \\
\hline
  0.5 & $10^4$ & 16 & 3.95543 & 36.36149 & 0.51906 & 1.41082 & 1.39596 \\
  0.5 & $10^4$ & 32 & 3.88987 & 36.31653 & 0.51882 & 1.40750 & 1.39267 \\
  0.5 & $10^4$ & 64 & 3.86941 & 36.31161 & 0.51879 & 1.40714 & 1.39232 \\
  0.5 & $10^4$ & 128 & 3.86383 & 36.31105 & 0.51879 & 1.40710 & 1.39228 \\
  \rowcolor{Gray} \rule{0pt}{2.2ex}
   &  & extrapolated & 3.86173 & 36.31098 & 0.51879 & 1.40710 & 1.39227 \\
  0.5 & $10^4$ & King UM & 3.851 & 36.307 & 0.519 & 1.404 & 1.391 \\
  0.5 & $10^4$ & King VT & 3.86 & 36.4 & 0.5188 & 1.41 & 1.393 \\
  0.5 & $10^4$ & King CU & 3.86 & 36.3 & 0.5191 & 1.4103 & 1.3948 \\
\hline
  1 & $10^4$ & 16 & 2.60286 & 26.04904 & 0.50879 & 1.46096 & 1.40373 \\
  1 & $10^4$ & 32 & 2.57654 & 26.03180 & 0.50864 & 1.45907 & 1.40188 \\
  1 & $10^4$ & 64 & 2.56869 & 26.02986 & 0.50862 & 1.45886 & 1.40168 \\
  1 & $10^4$ & 128 & 2.56661 & 26.02964 & 0.50862 & 1.45883 & 1.40166 \\
  \rowcolor{Gray} \rule{0pt}{2.2ex}
   &  & extrapolated & 2.56586 & 26.02961 & 0.50862 & 1.45883 & 1.40165 \\
  1 & $10^4$ & King UM & 2.556 & 26.007 & 0.509 & 1.459 & 1.396 \\
  1 & $10^4$ & King VT & 2.57 & 26.1 & 0.5088 & 1.465 & 1.4 \\
  1 & $10^4$ & King CU & 2.57 & 26 & 0.5092 & 1.4651 & 1.4019 \\
\hline
  0.25 & $10^5$ & 16 & 9.85211 & 180.27625 & 0.53099 & 2.09904 & 2.09486 \\
  0.25 & $10^5$ & 32 & 9.55792 & 178.73582 & 0.53064 & 2.06534 & 2.06106 \\
  0.25 & $10^5$ & 64 & 9.35214 & 178.44234 & 0.53038 & 2.05898 & 2.05470 \\
  0.25 & $10^5$ & 128 & 9.28366 & 178.40932 & 0.53035 & 2.05826 & 2.05399 \\
  \rowcolor{Gray} \rule{0pt}{2.2ex}
   &  & extrapolated & 9.24951 & 178.40514 & 0.53034 & 2.05817 & 2.05390 \\
  0.25 & $10^5$ & King UM & 9.211 & 178.56 & 0.53 & 2.046 & 2.053 \\
  0.25 & $10^5$ & King VT & 9.26 & 180.2 & 0.5303 & 2.06 & 2.055 \\
  0.25 & $10^5$ & King CU & 9.23 & 178.6 & 0.5303 & 2.0597 & 2.0573 \\
\hline
  0.5 & $10^5$ & 16 & 8.09115 & 157.65389 & 0.54628 & 3.32727 & 3.30146 \\
  0.5 & $10^5$ & 32 & 7.84156 & 156.53900 & 0.54587 & 3.28255 & 3.25680 \\
  0.5 & $10^5$ & 64 & 7.69385 & 156.34222 & 0.54551 & 3.27460 & 3.24890 \\
  0.5 & $10^5$ & 128 & 7.64696 & 156.31949 & 0.54546 & 3.27368 & 3.24800 \\
  \rowcolor{Gray} \rule{0pt}{2.2ex}
   &  & extrapolated & 7.62514 & 156.31653 & 0.54545 & 3.27357 & 3.24788 \\
  0.5 & $10^5$ & King UM & 7.588 & 156.503 & 0.545 & 3.258 & 3.245 \\
  0.5 & $10^5$ & King VT & 7.63 & 157.93 & 0.5454 & 3.279 & 3.25 \\
  0.5 & $10^5$ & King CU & 7.61 & 156.5 & 0.5455 & 3.2779 & 3.2552 \\
\hline
  1 & $10^5$ & 16 & 4.07456 & 85.14137 & 0.52985 & 2.83063 & 2.77505 \\
  1 & $10^5$ & 32 & 3.97234 & 84.89394 & 0.52980 & 2.81369 & 2.75800 \\
  1 & $10^5$ & 64 & 3.93521 & 84.87353 & 0.52966 & 2.81193 & 2.75626 \\
  1 & $10^5$ & 128 & 3.92499 & 84.87099 & 0.52964 & 2.81173 & 2.75606 \\
  \rowcolor{Gray} \rule{0pt}{2.2ex}
   &  & extrapolated & 3.92110 & 84.87063 & 0.52964 & 2.81170 & 2.75604 \\
  1 & $10^5$ & King UM & 3.907 & 85.105 & 0.529 & 2.802 & 2.75 \\
  1 & $10^5$ & King VT & 3.92 & 86.08 & 0.5297 & 2.821 & 2.757 \\
  1 & $10^5$ & King CU & 3.92 & 85.1 & 0.5297 & 2.8278 & 2.7725 \\
\hline
  \end{tabular}

  \caption{Compressible results using the TALA formulation
    corresponding to the benchmark defined in \cite{KLVLZTTK10}. All
    other data as in Table~\ref{fig:king_full_results_ala}.} 
 \label{fig:king_full_results_tala}
\end{table*}

\end{document}